\documentclass[11pt]{iopart}

\usepackage{iopams}
\usepackage{amsfonts,bbm,amssymb}
\usepackage{setstack}
\usepackage{dsfont}
\usepackage{hyperref}

\def\bra#1{\mathinner{\langle{#1}|}}
\def\ket#1{\mathinner{|{#1}\rangle}}

\def\ul#1{\underline{#1}}

\def\hatm#1{\hat{\mathcal{#1}}}
\def\bb#1{\mathbf{#1}}
\def\adH{\,{\rm ad}_{\bb{H}(x)}}

\newcommand{\ii}{ {\rm i} }
\newcommand{\dd}{ {\rm d} }
\newcommand{\ZZ}{\mathbb{Z}}
\newcommand{\RaR}{\mathbb{R}}
\newcommand{\CC}{\mathbb{C}}
\newcommand{\y}{{\rm y}}
\newcommand{\x}{{\rm x}}
\newcommand{\z}{{\rm z}}

\newcommand{\mm}[1]{{\mathbf{#1}}}

\def\tr{{{\rm tr}}}

\def\one{\mathbbm{1}}

\newcommand{\binom}[2]{\left({#1 \atop #2}\right)}

\eqnobysec

\begin{document}

\title[Exterior integrability]
{Exterior integrability: Yang-Baxter form of nonequilibrium steady
state density operator}
\author{Toma\v{z} Prosen, Enej Ilievski}

\address{Department of physics, Facultry of Mathematics and Physics, University of Ljubljana,
Jadranska 19, SI-1000 Ljubljana, Slovenia}

\author{Vladislav Popkov}

\address{Max Planck Institut for Physics of Complex Systems, N\"othnitzer Str. 38, D-01187 Dresden, Germany, and\\
Dipartimento di Fisica e Astronomia, Universit\'a di Firenze, via
G. Sansone 1, I-50019 Sesto Fiorentino, Italy}

\date{\today}

\begin{abstract}
A new type of quantum transfer matrix, arising as a Cholesky factor for the steady state density matrix of
a dissipative Markovian process associated with the boundary-driven Lindblad equation for the isotropic spin-$1/2$ Heisenberg ($XXX$) chain,
is presented. The transfer matrix forms a commuting family of non-Hermitian operators depending on the spectral parameter which is essentially the strength of dissipative coupling at the boundaries. The intertwining of the corresponding Lax and monodromy matrices is performed by an infinitely dimensional Yang-Baxter $R$-matrix which we construct explicitly and which is essentially different from the standard $4\times 4$ $XXX$ $R$-matrix. We also discuss a possibility to construct Bethe Ansatz for the spectrum and eigenstates of the non-equilibrium steady state density operator.
Furthermore, we indicate the existence of a deformed $R$-matrix in the infinitely-dimensional auxiliary space for the anisotropic $XXZ$ spin-$1/2$ chain
which in general provides a sequence of new, possibly quasi-local, conserved quantities of the bulk $XXZ$ dynamics.
\end{abstract}

\maketitle

\section{Introduction}

The theory of integrable quantum systems in $1+1$ dimensions, the
so-called {\em quantum inverse scattering}, is a well developed
field of mathematical physics \cite{Skly,TTF83,F94,KBI93} which
pioneered important new algebraic structures in pure mathematics,
such as quantum groups and their representations. The fundamental
object in this theory is the $R$-matrix, a solution of the
celebrated Yang-Baxter equation \cite{B82}, which gives rise to
integrable Hamiltonians possessing infinite families of conserved
quantities. Furthermore, these techniques often lead to explicit
methods for diagonalizing the Hamiltonian, such as Algebraic Bethe
ansatz (ABA) \cite{Skly,F94,KBI93} or Baxter $Q$-operator
$\cite{BLMS10}$. More recently, the theory of integrable quantum
systems also found applications in classical non-equilibrium
physics, namely in solving markovian stochastic many-body
interacting systems such as asymmetric simple exclusion process
\cite{S01,BE07}. There has been even an attempt to develop a
non-equilibrium Bethe ansatz approach to quantum impurity problems
\cite{MA06}, nevertheless the practical feasibility of these
technique and its relation to general integrability structures
such as Yang-Baxter equations remains unclear.

However, very recently explicit results appeared for driven quantum many-body systems with a strong interaction, namely a closed {\em matrix product ansatz} (MPA) for non-equilibrium steady state (NESS) density-operator of the boundary-driven Lindblad equation \cite{P11a,P11b,P12a} of an anisotropic Heisenberg ($XXZ$) spin $1/2$ chain. Lindblad equation \cite{L76,GKS76} is the canonical model of continuous-time Markovian quantum dynamics.
This solution has been later interpreted in terms of infinite-dimensional representations of Lie algebra $\mathfrak{sl}(2)$, and its quantum-group deformation for the anisotropic spin interaction, and generalized to more general boundary dissipators/drivings \cite{KPS13}.
Remarkably, perturbative expansion of NESS in the dissipation strength gave rise to a novel $XXZ$ quasi-local conservation law \cite{P11a} which is unrelated to previously known local conserved quantities of the $XXZ$ chain \cite{GM95} derived from
the `standard' $XXZ$ transfer-matrix, and which has important consequences for understanding ballistic transport at high-temperatures \cite{IP13}.

In this paper we put these results \cite{P11a,P11b,P12a,KPS13} into the framework of the theory of integrable systems. Focusing mainly on the isotropic case ($XXX$ model), we rigorously construct an $R-$matrix satisfying Yang-Baxter in an infinitely
dimensional auxiliary space which carries irreducible infinitely-dimensional representation of $\mathfrak{sl}(2)$, so that the corresponding family of commuting transfer-matrices is given by the Cholesky factor of the unnormalized NESS density operator \cite{P11b}.
However, the commuting transfer matrix is given as the ground-state matrix element of the monodromy matrix, and not as its trace as in standard ABA, and is neither a Hermitian nor a diagonalizable operator, which is a manifestation of far-from-equilibrium character of the problem.
As the spectral parameter in our $R-$matrix comes from the boundary dissipative coupling we chose to call our formalism {\em the exterior integrability}.
One may also provide arguments for the existence of a deformed version of the infinite dimensional exterior $R-$matrix in the anisotropic ($XXZ$) case.
Two important immediate applications of the new $R-$matrix are proposed: (i) Construction of an infinite family of new almost-conserved \cite{IP13} quantities mutually in involution which include the one discussed in \cite{P11a} and which should shed further light on the understanding of
finite-temperature quantum transport problem \cite{SPA11,S12,HHB07}, and (ii) Construction of ABA for diagonalization of NESS density operator. We stress that even if the exterior integrability is defined with respect to particular integrable dissipative boundaries, it may produce interesting new results for bulk properties of the system in the thermodynamic limit, such as the quasi-local conserved quantities.

After defining the main concepts of non-equilibrium quantum integrability of the $XXX$ model in Section 2, we write explicit expression for the corresponding infinite $R-$matrix in Section 3 and rigorously prove that it satisfies the Yang-Baxter equation.
In Section 4 we describe some interesting properties of the $R$-matrix and the corresponding non-equilibrium monodromy matrix.
In Section 5 we discuss potential applications and extension to an anisotropic case, and conclude. Some technical aspects of our proofs are put into appendices.
While material presented in Sections 2, 3 and the appendices should be mathematically rigorous, further results discussed in Sections 4 and 5 are partly based on heuristic and empirical arguments.

\section{Exterior integrability of the nonequilibrium steady state}

We focus on the stationary Lindblad equation for the NESS density operator $\rho_\infty$
\begin{equation}
\ii[H_{XXX},\rho_\infty]=\varepsilon \hatm{D}(\rho_\infty)
\label{eq:lind}
\end{equation}
for the Heisenberg $XXX$ Hamiltonain of a chain of $n$ spins $1/2$
\begin{equation}
H_{XXX}=\sum_{j=1}^{n-1}\one_{2^{j-1}}\otimes h \otimes \one_{2^{n-j-1}},
\quad h=2\sigma^+\otimes \sigma^{-}
+2\sigma^{-}\otimes \sigma^{+} + \sigma^{\z}\otimes \sigma^{\z},\quad
\label{xxx}
\end{equation}
where $\sigma^{\pm}=\frac{1}{2}(\sigma^\x \pm \ii\sigma^\y)$, $\sigma^\z$, $\sigma^0=\one_2$ are standard Pauli matrices acting over a $2-$dimensional quantum spin space ${\cal H}_{\rm s}\simeq \CC^2$ and $\one_d$ is a $d$-dimensional unit matrix.
We chose the simplest solvable far-from-equilibrium dissipative driving \cite{P11b} with a pair of Lindblad jump operators with dissipation-driving strength $\varepsilon$
\begin{equation}
\hatm{D}(\rho)=\sum_{k=1}^2 \bigl( 2L_{k}\rho L_{k}^{\dagger}-\{L_{k}^{\dagger}L_{k},\rho\} \bigr),
\; L_{1}=\sigma^{+} \otimes \one_{2^{n-1}},\; L_{2}= \one_{2^{n-1}}\otimes \sigma^{-}.
\end{equation}
As it has been shown in \cite{P11b}, the unique NESS density operator can be written explicitly in the {\em Cholesky factorized} form
\begin{equation}
\tilde{\rho}_{\infty}=S(\lambda)S^{\dagger}(\lambda),\qquad \rho_{\infty}=\frac{\tilde{\rho}_{\infty}}{\tr{\rho_{\infty}}}.
\label{Choleskyform}
\end{equation}
where the operator $S(\lambda)$ admits an elegant representation in terms of MPA:
\begin{equation}
S(\lambda)=\sum_{s_1,\ldots,s_n\in\{+,0,-\}} \bra{0}\mm{A}_{s_1}(\lambda)\cdots \mm{A}_{s_n}(\lambda)\ket{0} \sigma^{s_1}\otimes \cdots \otimes \sigma^{s_n}
\label{eq:sol}
\end{equation}
where
\begin{eqnarray}
 \bb{A}_{0}(\lambda) &=& \sum_{k=0}^\infty a^{0}_{k}(\lambda)\ket{k}\bra{k},\nonumber\\
 \bb{A}_{+}(\lambda) &=& \sum_{k=0}^\infty a^{+}_{k}(\lambda)\ket{k}\bra{k+1},\label{MPOtensors}\\
\bb{A}_{-}(\lambda) &=& \sum_{k=0}^\infty a^{-}_{k}(\lambda)\ket{k+1}\bra{k},\nonumber
\end{eqnarray}
is family of tridiagonal matrix operators acting on an {\em infinitely-dimensional} auxiliary Hilbert space ${\cal H}_{\rm a}$ with a canonical basis $\{\ket{0},\ket{1},\ket{2},\ldots\}$.
In fact, the consistency of solution (\ref{eq:sol}) with the defining equation (\ref{eq:lind}) requires that the matrix operators span an infinitely dimensional irreducible representation of $\mathfrak{sl}(2)$ algebra
\begin{equation}
[\bb{A}_{+}(\lambda),\bb{A}_{-}(\lambda)]=-2\bb{A}_{0}(\lambda),\qquad [\bb{A}_{0}(\lambda),\bb{A}_{\pm}(\lambda)]=\pm \bb{A}_{\pm}(\lambda)
\end{equation}
which may be -- up to unitary transformations -- uniquely chosen as\footnote{Note that the representation of Ref.\cite{P11b}, specialized to the isotropic case, is, up to a constant, unitarily equivalent to (\ref{MPOamplitudes}).}
\begin{equation}
a^{0}_{k}(\lambda)=\lambda-k,\qquad a^{+}_{k}(\lambda)= k-2\lambda,\qquad a^{-}_{k}(\lambda)= k+1,\qquad \lambda\in \mathbb{C},\;
\label{MPOamplitudes}
\end{equation}
with the complex representation parameter $\lambda$ being fixed by the boundary dissipation strength
\begin{equation}
\lambda =\frac{2\rmi}{\varepsilon}.
\end{equation}
Defining a $\lambda$-dependent linear operator from ${\rm End}({\cal H}_{\rm s}\otimes{\cal H}_{\rm a})$
\begin{equation}
\bb{L}(\lambda)=\sigma^{0}\otimes \bb{A}_{0}(\lambda) + \sigma^{+}\otimes \bb{A}_{+}(\lambda) + \sigma^{-}\otimes \bb{A}_{-}(\lambda)
=\pmatrix{\bb{A}_0(\lambda) & \bb{A}_+(\lambda) \cr
\bb{A}_-(\lambda) & \bb{A}_0(\lambda)},
\label{Lax}
\end{equation}
the Cholesky factor can be expressed even more elegantly \cite{KPS13}
\begin{equation}
S(\lambda)=\bra{0}\bb{L}(\lambda)\otimes_{\rm s} \bb{L}(\lambda)\otimes_{\rm s} \ldots \otimes_{\rm s} \bb{L}(\lambda)\ket{0}=
\bra{0}\bb{L}(\lambda)^{\otimes_{\rm s}n}\ket{0}.
\label{MPA}
\end{equation}
Here and below we use the following compact and unambiguous notational convention. For operator-valued matrices,
we use a symbol $\otimes_{\rm s}$ as a {\em partial tensor product}, namely it implies a tensor product with respect to the quantum spin space ${\cal H}_{\rm s}$ and an ordinary operator/matrix product with respect to the auxiliary space
${\cal H}_{\rm a}$. Analogously, $\otimes_{\rm a}$ will denote a tensor product with respect to ${\cal H}_{\rm a}$, and a matrix product in ${\cal H}_{\rm s}$.
For example, for $\mm{X}\in{\rm End}({\cal H}_{\rm s}^{\otimes j}\otimes {\cal H}_{\rm a}^{\otimes k})$, $\mm{Y}\in{\rm End}({\cal H}_{\rm s}^{\otimes l}\otimes {\cal H}_{\rm a}^{\otimes m})$,
$\mm{X}\otimes_{\rm s}\mm{Y}\in {\rm End}({\cal H}_{\rm s}^{\otimes (j+l)}\otimes {\cal H}_{\rm a}^{\otimes k})$ making sense if $k=m$, and
$\mm{X}\otimes_{\rm a}\mm{Y}\in {\rm End}({\cal H}_{\rm s}^{\otimes j}\otimes {\cal H}_{\rm a}^{\otimes (k+m)})$ making sense if $j=l$.
To emphasize the exterior integrability concepts we shall write in bold all symbols which are {\em not scalars} with respect to auxiliary space ${\cal H}_{\rm a}$.

The key step of this work is to recognize that $\bb{L}(\lambda)$ can be interpreted as the Lax matrix (the so-called $L-$matrix) matrix of a novel integrable system.
This is founded on a simple empirical observation, namely that the Cholesky factors commute for arbitrary complex values of the representation/dissipation parameters
\begin{equation}
[S(\lambda),S(\mu)] = 0, \quad \forall \lambda,\mu\in\CC.
\label{commut}
\end{equation}
This observation can be understood as a consequence of existence of an $R-$matrix\footnote{We follow nomenclature of Ref.~\cite{KBI93} here. Often in literature the term $R$-matrix is reserved for an operator $\mm{P}\mm{R}$ where $\mm{P}$ is
permutation operator which swaps the auxiliary spaces.} $\mm{R}(\lambda,\mu) \in {\rm End}({\cal H}_{\rm a}\otimes {\cal H}_{\rm a})$ for {\em almost any} $\lambda,\mu\in\CC$, to be shown in Section 3, which satisfies the so-called $RLL$ (or local intertwining) relation:
\begin{equation}
\bb{R}(\lambda,\mu)\left(\bb{L}(\lambda)\otimes_{\rm a} \bb{L}(\mu)\right)=\left(\bb{L}(\mu)\otimes_{\rm a} \bb{L}(\lambda)\right)\bb{R}(\lambda,\mu),
\label{RLL}
\end{equation}
Following the procedure of ABA \cite{KBI93} the local intertwining relation immediately implies intertwining for a product of the so-called {\em monodromy} matrices
$\bb{T}(\lambda)\in{\rm End}({\cal H}^{\otimes n}_{\rm s}\otimes {\cal H}_{\rm a})$:
\begin{equation}
\bb{T}(\lambda)=\bb{L}(\lambda)\otimes_{\rm s} \bb{L}(\lambda)\otimes_{\rm s} \ldots \otimes_{\rm s} \bb{L}(\lambda) = \bb{L}(\lambda)^{\otimes_{\rm s} n},
\label{T}
\end{equation}
namely
\begin{equation}
\bb{R}(\lambda,\mu)\left(\bb{T}(\lambda)\otimes_{\rm a} \bb{T}(\mu)\right)=\left(\bb{T}(\mu)\otimes_{\rm a} \bb{T}(\lambda)\right)\bb{R}(\lambda,\mu).
\label{RTT}
\end{equation}
Indeed, Eq. (\ref{RLL}) implies Eq. (\ref{RTT}) after noticing that, due to associativity of matrix multiplication:
\begin{equation}
\bb{T}(\lambda)\otimes_{\rm a} \bb{T}(\mu)
= (\bb{L}(\lambda)^{\otimes_{\rm s} n}) \otimes_{\rm a}
(\bb{L}(\mu)^{\otimes_{\rm s} n}) =
(\bb{L}(\lambda) \otimes_{\rm a} \bb{L}(\mu))^{\otimes_{\rm s} n}
\end{equation}

Unlike in the standard formalism of ABA where the auxiliary space is finite dimensional and the concept of a transfer matrix is usually associated to the partial trace of monodromy matrix with respect to the auxiliary space, we define here the auxiliary ground state expectation $\bra{0}\mm{T}(\lambda)\ket{0}=S(\lambda)$ as the transfer matrix.
In order to establish the commutativity of the transfer matrix we also require, besides the $RTT$ relations (\ref{RTT}), the $R$-matrix
to satisfy additional boundary conditions
\begin{equation}
\bra{0,0}\bb{R}(\lambda,\mu)=\bra{0,0},\qquad \bb{R}(\lambda,\mu)\ket{0,0}=\ket{0,0},
\label{Rbc}
\end{equation}
where $\ket{k,l}:=\ket{k}\otimes\ket{l}, \bra{k,l}:=\bra{k}\otimes\bra{l}$. Eq. (\ref{commut}) then follows straightforwardly, after writing the transfer-matrix product in ${\cal H}_{\rm a}\otimes {\cal H}_{\rm a}$,
$S(\lambda)S(\mu) = \bra{0,0}\bb{T}(\lambda)\otimes_{\rm a} \bb{T}(\mu)\ket{0,0}$:
\begin{eqnarray}
S(\lambda)S(\mu)&=\bra{0,0}\bb{R}(\lambda,\mu)\left(\bb{T}(\lambda)\otimes_{\rm a} \bb{T}(\mu)\right)\ket{0,0}\nonumber \\
&=\bra{0,0}\left(\bb{T}(\mu)\otimes_{\rm a} \bb{T}(\lambda)\bb{R}(\lambda,\mu)\right)\ket{0,0}=S(\mu)S(\lambda).
\label{involution}
\end{eqnarray}
Despite the boundary condition (\ref{Rbc}) may seem a-priori unjustified at the moment, we shall show further on, that
such a property naturally follows from the so-called ice-rule property of the $R$-matrix.

It is perhaps remarkable that the transfer matrix of our problem $S(\lambda)$ is non-Hermitian, non-normal, and even non-diagonalizable operator.
Using the MPA form (\ref{MPA}) we can write its matrix elements in the spin basis $\{ \ket{\ul{\nu}} = \ket{\nu_1,\ldots,\nu_n}; \nu_j \in\{0,1\} \}$ of ${\cal H}_{\rm s}^{\otimes n}$, $\sigma^\z\ket{\nu} = (-1)^\nu \ket{\nu}$,
as
\begin{equation}
\bra{\ul{\nu}'}S(\lambda)\ket{\ul{\nu}} =
\bra{0}\mm{A}_{\nu_1-\nu_1'}(\lambda)\mm{A}_{\nu_2-\nu_2'}(\lambda)\cdots \mm{A}_{\nu_n-\nu_n'}(\lambda)\ket{0}
\end{equation}
so that the rules $\bra{0}\mm{A}_0 = \lambda \bra{0}$ and $\bra{0}\mm{A}_- = 0$, following from representation (\ref{MPOtensors}), imply the matrix of $S(\lambda)$ to be {\em upper triangular},
\begin{equation}
\bra{\ul{\nu}'}S(\lambda)\ket{\ul{\nu}}=0,\quad{\rm if}\quad \sum_{j=1}^n \nu'_j 2^{n-j} > \sum_{j=1}^n \nu_j 2^{n-j},
\end{equation}
and having a constant diagonal
\begin{equation}
\bra{\ul{\nu}}S(\lambda)\ket{\ul{\nu}}=\lambda^n.
\end{equation}
Consequently, all eigenvalues of $S(\lambda)$ are equal to $\lambda^n$, but since $S(\lambda)$ is {\em not} a multiple of the identity operator it must have a non-trivial Jordan decomposition, i.e. it must be non-diagonalizable.

Similarly, we can write the quantum space matrix elements of the general monodromy matrix elements
\begin{equation}
T^{k'}_k(\lambda) := \bra{k'}\mm{T}(\lambda)\ket{k}
\label{meT}
\end{equation}
following the expression (\ref{T}) in terms of MPA
\begin{equation}
\bra{\ul{\nu}'}T^{k'}_k(\lambda)\ket{\ul{\nu}} = \bra{k'}\mm{A}_{\nu_1-\nu_1'}(\lambda)\mm{A}_{\nu_2-\nu_2'}(\lambda)\cdots \mm{A}_{\nu_n-\nu_n'}(\lambda)\ket{k}.
\label{Tmatrix}
\end{equation}
Tridiagonality of operators (\ref{MPOtensors}) immediately implies a magnetization selection rule, namely (\ref{Tmatrix}) vanishes unless
\begin{equation}
k'-k+ \sum_{j=1}^n \nu_j-\nu'_j = 0.
\label{selrule}
\end{equation}
This in turn implies that $T^{k'}_k(\lambda)$ changes the $z-$component of magnetization by $2(k'-k)$,
\begin{equation}
[M,T^{k'}_k(\lambda)]
=2(k'-k)T^{k'}_k(\lambda),
\end{equation}
writing magnetization operator as $M:=\sum_{j=1}^n \one_{2^{j-1}}\otimes \sigma^\z \otimes \one_{2^{n-j}}$.

\subsection{Ice-rule -- the particle conservation law}

Let us write out the $R$-matrix in components
\begin{equation}
\bb{R}(\lambda,\mu)=\sum_{k,k'=0}^{\infty}\sum_{l,l'=0}^{\infty}R^{kk'}_{ll'}(\lambda,\mu) \ket{k,k'}\bra{l,l'}.
\end{equation}
We will show in the following section that the exterior $R-$matrix of the $XXX$ model (and also for a more general $XXZ$ model, see subsection \ref{xxzsection})
obeys a selection rule, namely $R^{kk'}_{ll'}(\lambda,\mu) \ne 0$ only if $k+k'=l+l'$.
This can be interpreted as a particular particle conservation (global $U(1)$) symmetry of the $R-$matrix, meaning that it should commute with the particle number operator
\begin{eqnarray}
&& [\bb{R}(\lambda,\mu),\bb{N}]=0, \\
&& \bb{N} =-(\bb{A}_{0}(0)\otimes \mathds{1}+\mathds{1}\otimes \bb{A}_{0}(0))=\bigoplus_{\alpha=0}^{\infty}
\alpha\mathds{1}_{\alpha+1}.
\label{N}
\end{eqnarray}
Consequently, one can interpret the $R$-matrix as a particle-number conserving scattering matrix of a system of auxiliary quasi-particles.
Decomposition (\ref{N}) suggests a natural splitting of a tensor product of two copies of auxiliary space into a direct sum of eigenspaces of $\bb{N}$
\begin{equation}
{\cal H}_{\rm a}\otimes{\cal H}_{\rm a} = \bigoplus_{\alpha=0}^\infty {\cal H}^{(\alpha)}_{\rm a}.
\end{equation}
As we see, there are $\alpha+1$ states $\ket{k,\alpha-k}$ within each sector ${\cal H}^{(\alpha)}_{\rm a}$.
Therefore,
for any $\bb{X}\in {\rm End}({\cal H}_{\rm a}\otimes{\cal H}_{\rm a})$ which commutes with $\bb{N}$, $[\bb{X},\bb{N}]=0$,
we shall denote with upper-bracketed index $\alpha$ an $(\alpha+1)\times(\alpha+1)$-matrix component
of decomposition $\bb{X}=\bigoplus_{\alpha=0}^{\infty} \bb{X}^{(\alpha)}$.
For example, we shall often write the $R-$matrix in the so-called ice-rule form
\begin{equation}
\bb{R}(\lambda,\mu) = \sum_{\alpha=0}^\infty \sum_{k,l=0}^{\alpha} R^{(\alpha)}_{k,l}(\lambda,\mu) \ket{k,\alpha-k}\bra{l,\alpha-l} = \bigoplus_{\alpha=0}^\infty \bb{R}^{(\alpha)}(\lambda,\mu).
\label{icerule}
\end{equation}
As elements of ${\cal H}^{(0)}_{{\rm a}}$ are {\em scalars}, any $R-$matrix satisfying the ice-rule (\ref{icerule}) should trivially obey
the boundary condition (\ref{Rbc}).

\section{Exterior $R$-matrix}

Here we shall write out and prove our main result, an explicit form of the infinitely dimensional exterior $R-$matrix which satisfies the defining $RLL$ relations (\ref{RLL}).
\\\\
{\bf Theorem.} A solution of the $RLL$ (\ref{RLL}) relation for Lax operator \eref{Lax} reads
\begin{equation}
\bb{R}(x+{\textstyle\frac{1}{2}}y,x-{\textstyle\frac{1}{2}}y)=\exp{(y\bb{H}(x))},
\label{expform}
\label{theorem}
\end{equation}
for any $x\in \CC\setminus \frac{1}{2}\ZZ_+$, $y\in \CC$.
The generator $\bb{H}(x)$ admits a block decomposition according to the ice-rule,
\begin{equation}
\bb{H}(x)=\bigoplus_{\alpha}\bb{H}^{(\alpha)}(x),
\label{eq:Hdec}
\end{equation}
with explicit form of the matrix elements
\begin{eqnarray}
H^{(\alpha)}_{k,l}(x)=\frac{(-1)^{k-1}}{2}{k \choose l}\sum_{m=l}^{k-1}(-1)^{m}{k-l-1 \choose m-l}f_{m}(x),\quad k\geq l+1,\nonumber\\
H^{(\alpha)}_{k,k}(x)=-\frac{1}{2}\sum_{m=k}^{\alpha-k-1}f_{m}(x),\qquad 2k\leq \alpha,  \label{firstform}\\
H^{(\alpha)}_{\alpha-k,\alpha-l}(x)=-H^{(\alpha)}_{k,l}(x),\nonumber
\end{eqnarray}
where we introduced simple-pole functions $f_{m}(x):=(x-m/2)^{-1}$.

\medskip
\noindent
{\bf Proof.}
We start by using (\ref{theorem}) as an ansatz and reparametrize the $RLL$ relation (\ref{RLL}) in the {\em center-of-mass} and
{\em displacement} spectral parameters,
\begin{equation}
x=(\lambda+\mu)/2,\qquad y=\lambda-\mu,
\end{equation}
namely
\begin{equation}
\fl \exp(y\bb{H}(x))\left(\bb{L}\left(x\!+\!{\textstyle\frac{1}{2}}y\right)\otimes_{\rm a} \bb{L}\left(x\!-\!{\textstyle\frac{1}{2}}y\right)\right)=
\left(\bb{L}\left(x\!-\!{\textstyle\frac{1}{2}}y\right)\otimes_{\rm a} \bb{L}\left(x\!+\!{\textstyle\frac{1}{2}}y\right)\right)\exp(y\bb{H}(x)),
\label{RLLmain}
\end{equation}
yielding the form where non-trivial dependence enters through the generator $\bb{H}(x)$, in a way which resembles a Lie group structure.
Furthermore, we employ the fact that the Lax matrix $\bb{L}(x)$ has a simple \textit{linear dependence} on the spectral parameter \begin{equation}
\bb{L}(\lambda)=\bb{L}_0+\lambda \bb{L}^{\prime},\quad \bb{L}_0:=\bb{L}(0),\;\bb{L}':=\frac{\dd}{\dd x}\bb{L}(x),
\end{equation}
whence
\begin{eqnarray}
\bLambda(x,y)&:=
\bb{L}\left(x+{\textstyle\frac{1}{2}}y\right)\otimes_{\rm a} \bb{L}\left(x-{\textstyle\frac{1}{2}}y\right) \nonumber\\
&=\bb{L}(x)\otimes_{\rm a} \bb{L}(x)\nonumber -\frac{y}{2}\left(\bb{L}(x)\otimes_{\rm a} \bb{L}^{\prime}-\bb{L}^{\prime}\otimes_{\rm a} \bb{L}(x)\right)-\frac{y^2}{4}\left(\bb{L}^{\prime}\otimes_{\rm a} \bb{L}^{\prime}\right)\nonumber \\
&=:\bLambda_{0}(x)-\frac{y}{2}\bLambda_{1}-\frac{y^2}{4}\bLambda_{2}. \label{eq:Lambda}
\end{eqnarray}
At this point we emphasize that the whole $x$-dependence is absorbed into zero-th degree component $\bLambda_0(x)$, whereas $\bLambda_{1,2}$
are matrices with constant ($x$ independent) elements.
In particular, $\bb{\Lambda}_1 = \bb{L}(x)\otimes\bb{L}' -  \bb{L}'\otimes\bb{L}(x) =  \bb{L}_0\otimes\bb{L}' -  \bb{L}'\otimes\bb{L}_0$.
Writing the Weyl basis of ${\rm End}({\cal H}_{\rm s})$ as $E^{\nu,\nu'} = \ket{\nu}\bra{\nu'}$ and expressing
\begin{eqnarray}
\bb{L}^{\prime}&=(E^{00}+E^{11})\otimes \mathds{1}- E^{01}\otimes\bb{B},\nonumber \\
\bb{B}&:=-\frac{\dd}{\dd x}\bb{A}_{+}(x)=2\sum_{k}\ket{k}\bra{k+1},
\end{eqnarray}
we can write the three orders $\bb{\Lambda}_{0,1,2}$ (\ref{eq:Lambda}) as operators over  $\mathcal{H}_{\rm s}\otimes \mathcal{H}_{\rm a}\otimes {\cal H}_{\rm a}$
factoring out the components in the physical space
\begin{eqnarray}
\!\!\!\!\!\!\!
\bLambda_{0}(x)&=E^{00}\otimes \left(\bb{A}_{0}(x)\otimes \bb{A}_{0}(x)+\bb{A}_{+}(x)\otimes \bb{A}_{-}\right)\nonumber \\
&+E^{11}\otimes \left(\bb{A}_{0}(x)\otimes \bb{A}_{0}(x)+\bb{A}_{-}\otimes \bb{A}_{+}(x)\right)\nonumber \\
&+E^{01}\otimes \left(\bb{A}_{0}(x)\otimes \bb{A}_{+}(x)+\bb{A}_{+}(x)\otimes \bb{A}_{0}(x)\right)\nonumber \\
&+E^{10}\otimes \left(\bb{A}_{0}(x)\otimes \bb{A}_{-}+\bb{A}_{-}\otimes \bb{A}_{0}(x)\right), \label{Lam0}\\
\!\!\!\!\!\!\!
\bLambda_{1}&=E^{00}\otimes\left(\bb{A}_{0}(0)\otimes \mathds{1}-\mathds{1}\otimes \bb{A}_{0}(0)+
\bb{B}\otimes \bb{A}_{-}\right)\nonumber \\
&+E^{11}\otimes\left(\bb{A}_{0}(0)\otimes \mathds{1}-\mathds{1}\otimes \bb{A}_{0}(0)-\bb{A}_{-}\otimes \bb{B}\right)\nonumber \\
&+E^{01}\otimes\left(\bb{A}_{+}(0)\otimes \mathds{1}-\mathds{1}\otimes \bb{A}_{+}(0)+
\bb{B}\otimes \bb{A}_{0}(0)-\bb{A}_{0}(0)\otimes \bb{B}\right)\nonumber \\
&+E^{10}\otimes(\bb{A}_{-}\otimes \mathds{1}-\mathds{1}\otimes \bb{A}_{-}),\label{Lam1}\\
\!\!\!\!\!\!\!
\bLambda_{2}&=(E^{00}+E^{11})\otimes (\mathds{1}\otimes \mathds{1})-
E^{01}\otimes(\bb{B}\otimes \mathds{1}+\mathds{1}\otimes \bb{B}). \label{Lam2}
\label{lambdas}
\end{eqnarray}

After inserting proposed ansatz for the solution \eref{theorem}, we shall expand \eref{RLLmain} in terms of \textit{nested commutators} --
(i) we multiply \eref{RLLmain} by the operator $\exp{(-\frac{y}{2}\bb{H}(x))}$ from the left and from the right, and (ii) we utilize the defining Lie-group identity
$\exp({\,{\rm ad}_X})Y=e^{X}Ye^{-X}$, where ${\rm ad}_X:=[X,\bullet]$, which brings (\ref{RLLmain}) to an equivalent form
\begin{equation}
\exp{\left({\textstyle\frac{1}{2}}y\adH\right)}\bLambda(x,y)-\exp{\left(-{\textstyle\frac{1}{2}}y\adH\right)}\bLambda(x,-y)=0,
\end{equation}
or
\begin{equation}
\sinh{\left({\textstyle\frac{1}{2}}y\adH\right)}\left(\bLambda_0(x)-{\textstyle\frac{1}{4}}y^2\bLambda_2\right)-{\textstyle{\frac{1}{2}}}y\cosh{\left({\textstyle\frac{1}{2}}y\adH\right)}\bLambda_1 = 0.
\end{equation}
Expanding the hyperbolic functions we obtain a power series in $y$, which always exists in terms of finite matrix exponentials due to decomposition (\ref{eq:Hdec}).
Clearly, since the expression above is an odd function in $y$, we find only {\em odd} orders nonvanishing.
In the first order in $y$ we have
\begin{equation}
\adH\bLambda_{0}(x)=\bLambda_{1},
\label{HLLrelation}
\end{equation}
while for an arbitrary odd order $y^{2l+1}$ with $l \ge 1$:
\begin{equation}
\adH^{2l+1}\bLambda_{0}(x)-(2l+1)\adH^{2l}\bLambda_{1}-2l(2l+1)\adH^{2l-1}\bLambda_{2}=0.
\label{adjointexpansion}
\end{equation}
The relation (\ref{HLLrelation}) is in fact an infinitesimal $RLL$ relation for $y\to\dd y$ and will be in the following referred to as $HLL$ relation.

Next we show that an infinite sequence of operator equations (\ref{adjointexpansion}) can be in fact reduced to only two equations.
The first one is just the {\em third} order [(\ref{adjointexpansion}) for $l=1$] after substituting $\adH \bLambda_0(x)$ from $HLL$ relation (\ref{HLLrelation}):
\begin{equation}
\adH^{2}\bLambda_{1}=-3\adH\bLambda_{2},
\label{HLL2nd}
\end{equation}
Then we subsequently use (\ref{HLLrelation}) and (\ref{HLL2nd}) to eliminate $\bLambda_0(x)$ and $\bLambda_1$ from the sequence (\ref{adjointexpansion})
for any $l>1$, arriving at $\adH^{2l-1} \bLambda_2=0$, for which a {\em sufficient condition} is
\begin{equation}
\adH^2 \bLambda_2 = 0.
\label{HLL3rd}
\end{equation}

We have thus shown that three simple $y-$independent equations, namely (\ref{HLLrelation}), and a pair (\ref{HLL2nd},\ref{HLL3rd})
imply validity of Eq. (\ref{adjointexpansion}) for any $l$, and consequently of the full $RLL$ relation for any pair of spectral parameters $x,y$ for which $\bb{H}(x)$ exists,
i.e.  $x\in \CC\setminus \frac{1}{2}\ZZ_+$, $y\in \CC$.

The remainder of the proof is thus to verify identities (\ref{HLLrelation},\ref{HLL2nd},\ref{HLL3rd}) which we formulate in two lemmas below.

\subsection{The $HLL$ relation}
\label{lemma1}

{\bf Lemma 1.}
The generator of the $R$-matrix $\bb{H}(x)$ (\ref{firstform}) solves the infinitesimal $RLL$ relation (\ref{HLLrelation}):
\begin{equation}
[\bb{H}(x),\bLambda_0(x)]=\bLambda_1,
\label{HLLlemma}
\end{equation}
for any $x\in \CC\setminus \frac{1}{2}\ZZ_+$ for which it is defined.

\medskip
\noindent
{\bf Proof.}
Let us define a permutation map -- homomorphism -- over ${\rm End}(\mathcal{H}_{\rm a}\otimes \mathcal{H}_{\rm a})$
\begin{equation}
\pi_{\rm a}(\bb{X})=\bb{P}\bb{X}\bb{P}^{-1} = \bb{P}\bb{X}\bb{P},
\label{pia}
\end{equation}
where $\bb{P}$ is a permutation operator over ${\cal H}_{\rm a}\otimes {\cal H}_{\rm a}$, acting as
\begin{equation}
\bb{P}\ket{k,l} = \ket{l,k},\quad k,l\in\ZZ^+.
\end{equation}
Permutation operator conserves the number of auxiliary excitations, hence it satisfies the ice rule
\begin{equation}
\bb{P}=\bigoplus_{\alpha}\bb{P}^{(\alpha)},\qquad P^{(\alpha)}_{k,l} = \delta_{k+l,\alpha}.
\label{P}
\end{equation}
We may write shortly $\pi_{\rm a}(\bb{a}\otimes \bb{b})=\bb{b}\otimes \bb{a}$.
Then we define another map $\pi_{\rm s}$ over operators in the quantum spin space ${\rm End}(\mathcal{H}_s)$, by
\begin{equation}
\pi_{\rm s}(E^{\nu\nu'}) = E^{1-\nu',1-\nu},
\end{equation}
or equivalently, $\pi_{\rm s}(\sigma^0) = \sigma^0$, $\pi_{\rm s}(\sigma^\pm) =\sigma^\pm$, $\pi_{\rm s}(\sigma^\z) = -\sigma^\z$,
so the {\em full parity map} over ${\cal H}_{\rm s}\otimes {\cal H}_{\rm a} \otimes {\cal H}_{\rm a}$ is defined as
\begin{equation}
\pi = \pi_{\rm s} \otimes \pi_{\rm a}.
\label{pi}
\end{equation}
It is important to note that the operators $\bLambda_{0,1,2}$ and the
generator $\bb{H}(x)$ are eigenoperators of the parity map, i.e. they have well defined parities [see (\ref{Lam0},\ref{Lam1},\ref{Lam2})]:
\begin{eqnarray}
\pi(\bLambda_{k})&=(-1)^{k}\bLambda_{k},\quad k=0,1,2, \qquad \pi_{\rm a}(\bb{H})=-\bb{H}.
\label{peigenvalues}
\end{eqnarray}
Notice that $\bb{H}(x)$ operates trivially (i.e. as a scalar) in the physical space ${\cal H}_{\rm s}$.
The whole expression (\ref{HLLlemma}) is then an eigenoperator of $\pi$ with eigenvalue $-1$,
\begin{equation}
[\bb{H}(x),\bLambda_0(x)]-\bLambda_1 + \pi\left([\bb{H}(x),\bLambda_0(x)]-\bLambda_1\right)=0.
\label{HpiH}
\end{equation}

Let us now introduce the components in the quantum spin space, either in Weyl or Pauli basis,
$\bLambda^s_k,\bLambda^{\nu\nu'}_k \in {\rm End}({\cal H}_{\rm a}\otimes{\cal H}_{\rm a})$, namely
\begin{equation}
\bLambda_{k}(x)=\sum_{\nu,\nu'=0}^1 E^{\nu\nu'}\otimes \bLambda^{\nu\nu'}_{k}(x) = \sum_{s\in\{0,+,-,\z\}} \sigma^s \otimes \bLambda^{s}_k(x).
\end{equation}
The identity (\ref{HLLlemma}) to be proven then writes
\begin{equation}
\sum_{\nu,\nu'=0}^1 E^{\nu\nu'} \otimes ([\bb{H}(x),\bLambda^{\nu\nu'}_0(x)]-\bLambda^{\nu\nu'}_1) = 0.
\label{HLLcomp}
\end{equation}
whereas the symmetry relation (\ref{HpiH}), noting (\ref{peigenvalues}), can be rewritten as
\begin{equation}
(E^{00}-E^{11}) \otimes \left(([\bb{H}(x),\bLambda^{00}_0(x)]\!-\!\bLambda^{00}_1) - ([\bb{H}(x),\bLambda^{11}_0(x)]\!-\!\bLambda^{11}_1)\right)=0.\; \label{HpiH2}
\end{equation}
This means that out of four equations in ${\rm End}({\cal H}_{\rm a}\otimes{\cal H}_{\rm a})$, implied by (\ref{HLLcomp}), only three are independent, say the components $00$, $01=+$ and $10=-$.

Furthermore, we apply the $\alpha-$ decomposition of the $\bLambda_k$ operators
\begin{equation}
\bLambda^{s}_k = \bigoplus_{\alpha=0}^{\infty} \bLambda^{(\alpha)s}_k,
\end{equation}
where $\bLambda^{(\alpha){0,\z}}_k=\bLambda^{(\alpha)00}_k\pm \bLambda^{(\alpha)11}_k \in {\rm End}({\cal H}^{(\alpha)}_{\rm a})$ are $(\alpha+1)\times (\alpha+1)$ matrices,
while $\bLambda^{(\alpha)+}_k\in {\rm Lin}({\cal H}^{(\alpha)}_{\rm a},{\cal H}^{(\alpha+1)}_{\rm a})$, and
$\bLambda^{(\alpha)-}_k\in {\rm Lin}({\cal H}^{(\alpha+1)}_{\rm a},{\cal H}^{(\alpha)}_{\rm a})$ are
$(\alpha+1)\times (\alpha+2)$ and $(\alpha+2)\times (\alpha+1)$ matrices, respectively.
With a bit of patience one can derive explicit expressions from Eqs. (\ref{Lam0},\ref{Lam1},\ref{Lam2}), for the {\em constant} operators $\bLambda_1$ and $\bLambda_2$.
Using a compact notation for  a canonical basis of ${\cal H}^{(\alpha)}_{\rm a}$,
$\ket{k}\equiv \ket{k,\alpha-k}$, the only non-vanishing blocks/components are
\begin{eqnarray}
\bLambda^{(\alpha)0}_{1}&=\sum_{k=0}^{\alpha}2(\alpha-2k)\ket{k}\bra{k} \nonumber \\
&+\sum_{k=0}^{\alpha-1}\Big\{2(\alpha-k)\ket{k}\bra{k+1} - 2(k+1)\ket{k+1}\bra{k}\Big\}\label{L1} \\
\bLambda^{(\alpha)\z}_{1}&=\sum_{k=0}^{\alpha}2(\alpha-k)\ket{k}\bra{k+1} +\sum_{k=0}^{\alpha-1}2(k+1)\ket{k+1}\bra{k},\label{L2} \\
\bLambda^{(\alpha)+}_{1}&=\sum_{k=0}^{\alpha}\Big\{(3k-\alpha)\ket{k}\bra{k}+(3k-2\alpha)\ket{k}\bra{k+1}\Big\},\label{L3}\\
\bLambda^{(\alpha)-}_{1}&=\sum_{k=0}^{\alpha}\Big\{(k+1)\ket{k+1}\bra{k}+(k-\alpha-1)\ket{k}\bra{k}\Big\},\label{L4}\\
\bLambda^{(\alpha)+}_{2}&=\sum_{k=0}^{\alpha}\Big\{-2\ket{k}\bra{k}-2\ket{k}\bra{k+1}\Big\}.\label{L5}
\end{eqnarray}
The full set of finite matrix equations which remain to be verified then reads:
\begin{eqnarray}
[\bb{H}^{(\alpha)}(x),\bLambda^{(\alpha)00}_0(x)]=\bLambda^{(\alpha)00}_1,\\
\bb{H}^{(\alpha)}(x)\bLambda^{(\alpha)+}_0(x) - \bLambda^{(\alpha)+}_0(x)\bb{H}^{(\alpha+1)}(x) =\bLambda^{(\alpha)+}_1, \\
\bb{H}^{(\alpha+1)}(x)\bLambda^{(\alpha)-}_0(x) - \bLambda^{(\alpha)-}_0(x)\bb{H}^{(\alpha)}(x) =\bLambda^{(\alpha)-}_1.
\end{eqnarray}
For this one needs to show that for all equations residua at the possible poles, $x=p/2$, $p=0,\ldots,\alpha+1$, match as well as the remainders.
This is done in full detail in \ref{appHLL}.

\subsection{Master-symmetry of the $H$-matrix}
\label{lemma2}
{\bf Lemma 2.}
For any $x\in \CC\setminus \frac{1}{2}\ZZ_+$ for which $\bb{H}(x)$ (\ref{firstform}) is defined, it satisfies identities (\ref{HLL2nd},\ref{HLL3rd}):
\begin{eqnarray}
[\bb{H}(x),[\bb{H}(x),\bLambda_{1}]] &+& 3[\bb{H}(x),\bLambda_{2}]=0,\cr
[\bb{H}(x),[\bb{H}(x),\bLambda_{2}]] &=& 0. \label{lemma2}
\end{eqnarray}

\medskip
\noindent
{\bf Proof.}
Despite it might be tempting to attack the problem similarly as in the case of Lemma 1, a direct calculation reveals that one cannot
avoid binomial expressions with double summation involving linear combinations of quadratic terms (products of two binomial coefficients), which
are extraordinary difficult to deal with. Fortunately, as we demonstrate below, there exists an elegant algebraic recursive procedure
originating from an extra symmetry of the generator $\bb{H}(x)$. Since we are dealing with quadratic expressions in $\bb{H}(x)$,
whose blocks $\bb{H}^{(\alpha)}(x)$ are singular with \textit{one-dimensional null-space}, additional information about null-vectors of
$(\bb{H}^{(\alpha)})^{2}$ will be required as well.

Here we shall label quantum space components with the Pauli basis. According to the structure (\ref{Lam0},\ref{Lam1},\ref{Lam2}) the Eqs. (\ref{lemma2}) are equivalent
to five identities which can be cast in terms of operators over ${\cal H}_{\rm a}\otimes{\cal H}_{\rm a}$, $\{\bb{D}^{s}_{1},\bb{D}^{+}_{2}\}$, $s\in\{0,+,-,z\}$ (temporarily dropping
spectral parameter dependence for the rest of this proof),
\begin{eqnarray}
\bb{D}^{0}_{1}&:=[\bb{H},[\bb{H},\bLambda^{0}_{1}]]=0,\nonumber \\
\bb{D}^{\z}_{1}&:=[\bb{H},[\bb{H},\bLambda^{\z}_{1}]]=0,\nonumber \\
\bb{D}^{+}_{1}&:=[\bb{H},[\bb{H},\bLambda^{+}_{1}]]+3[\bb{H},\bLambda^{+}_{2}]=0,\nonumber\\
\bb{D}^{-}_{1}&:=[\bb{H},[\bb{H},\bLambda^{-}_{1}]]=0,\nonumber \\
\bb{D}^{+}_{2}&:=[\bb{H},[\bb{H},\bLambda^{+}_{2}]]=0.
\label{Doperators}
\end{eqnarray}
Consistently with our previous notation, we will place additional subscript index $\alpha$,
e.g. $\bLambda^{(\alpha)+}_{2}$, when referring to a single $\alpha$-subspace.

The key ingredient here is notification of a `conserved charge' $\bLambda^{-}_{1}$,
\begin{equation}
[\bb{H},\bLambda^{-}_{1}]=0,
\end{equation}
connecting two adjacent $\alpha$-blocks,
\begin{equation}
\bb{H}^{(\alpha+1)}\bLambda^{(\alpha)-}_{1}=\bLambda^{(\alpha)-}_{1}\bb{H}^{(\alpha)}.
\end{equation}
Because the above identity is to hold regardless of the value of $x$, we essentially have to prove
$[\bb{X}^{(\alpha)p},\bLambda^{(\alpha)-}_{1}]=0$ for all residue matrices (\ref{residuedecomposition}) for
$p=0,1,\ldots \alpha$, demanding to verify the identity
\begin{equation}
(l+1)X^{(\alpha+1)p}_{k,l+1}-kX^{(\alpha)p}_{k-1,l}-(\alpha-l+1)X^{(\alpha+1)p}_{k,l}+(\alpha-k+1)X^{(\alpha)p}_{k,l}=0,
\end{equation}
for every $k=0,1,\ldots,\alpha+1$, $l=0,1,\ldots \alpha$.
In fact, it is sufficient to consider the identity expressed in terms of tensors $\bb{Y}^{(\alpha)p}$ by virtue of parity symmetry (\ref{XP}),
\begin{equation}
(l+1)Y^{(\alpha+1)p}_{k,l+1}-kY^{(\alpha)p}_{k-1,l}-(\alpha-l+1)Y^{(\alpha+1)p}_{k,l}+(\alpha-k+1)Y^{(\alpha)p}_{k,l}=0,
\end{equation}
which reduces to trivially verifiable combinatorial identities upon substitution (\ref{Y}).

Next we state algebraic relations among $\bLambda^{(\alpha)s}_{1,2}$, which are straightforwardly verified using explicit representations
(\ref{L1}--\ref{L5}), namely
\begin{eqnarray}
\bLambda^{(\alpha+1)0}_{1}\bLambda^{(\alpha)-}_{1}&=\bLambda^{(\alpha)-}_{1}\bLambda^{(\alpha)0}_{1}.\nonumber \\
\bLambda^{(\alpha+1)\z}_{1}\bLambda^{(\alpha)-}_{1}&=\bLambda^{(\alpha)-}_{1}\bLambda^{(\alpha)\z}_{1}-2\bLambda^{(\alpha)-}_{1},\nonumber \\
\bLambda^{(\alpha)+}_{1}\bLambda^{(\alpha)-}_{1}&=\bLambda^{(\alpha-1)-}_{1}\bLambda^{(\alpha-1)+}_{1}+\bLambda^{(\alpha)\z}_{1},\nonumber \\
\bLambda^{(\alpha)+}_{2}\bLambda^{(\alpha)-}_{1}&=\bLambda^{(\alpha-1)-}_{1}\bLambda^{(\alpha-1)+}_{2}.
\label{lambdaidentities}
\end{eqnarray}

The idea is then to derive recursive relation in $\alpha$ for the operators $\{\bb{D}^{(\alpha)s}_{1},\bb{D}^{(\alpha)+}_{2}\}$, and use induction in $\alpha$,
along with the trivial initial conditions $\bb{D}^{(\alpha)s}_{1,2}=0$, for $\alpha=0,1$,
which are easy to check (e.g. by direct evaluation), to prove the identities \eref{Doperators}. Since all obtained recursions are treated in
analogous way, we choose to work out explicitly the one with $\bb{D}^{+}_{2}$. After expanding double commutator,
\begin{equation}
\bb{D}^{(\alpha)+}_{2}=(\bb{H}^{(\alpha)})^{2}\bLambda^{(\alpha)+}_{2}-2\bb{H}^{(\alpha)}\bLambda^{(\alpha)+}_{2}\bb{H}^{(\alpha+1)}+
\bLambda^{(\alpha)+}_{2}(\bb{H}^{(\alpha+1)})^{2},
\label{exampleD2}
\end{equation}
multiplying by $\bLambda^{(\alpha)-}_{1}$ from the right, using (i) $\bb{H}^{(\alpha+1)}\bLambda^{(\alpha)-}_{1}=\bLambda^{(\alpha)-}_{1}\bb{H}^{(\alpha)}$ and
(ii) $\bLambda^{(\alpha)+}_{2}\bLambda^{(\alpha)-}_{1}=\bLambda^{(\alpha-1)-}_{1}\bLambda^{(\alpha-1)+}_{2}$,
and commuting $\bLambda^{(\alpha)-}_{1}$ to the left, we
obtain
\begin{equation}
\bb{D}^{(\alpha)+}_{2}\bLambda^{(\alpha)-}_{1}=\bLambda^{(\alpha-1)-}_{1}\bb{D}^{(\alpha-1)+}_{2}.
\end{equation}
The relation we have just established is however not enough to conclude on vanishing of $\bb{D}^{(\alpha)+}_{2}$, provided $\bb{D}^{(\alpha-1)+}_{2}=0$.
The reason lies in the \textit{non-invertibility} of the rectangular matrix $\bLambda^{(\alpha)-}_{1}$. We can nonetheless cure this weakness if we show that
there exists an additional $(\alpha+2)$-dimensional vector $\bb{u}^{(\alpha+1)}$, \textit{linearly independent} of the column space of $\bLambda^{(\alpha)-}_{1}$, which is in a
null-space of $\bb{D}^{(\alpha)+}_{2}$,
\begin{equation}
\bb{D}^{(\alpha)+}_{2}\bb{u}^{(\alpha+1)}=0.
\label{nullspacevector}
\end{equation}
Notice that the remaining $\alpha+1$ columns of $\bLambda^{(\alpha)-}_{1}$ are indeed linearly independent (another simple calculation).

A crucial observation is, that for every $\alpha$-sector, there exist a unique pair of $(\alpha+1)$-dimensional null-vectors
$\bb{u}^{(\alpha)},\bb{v}^{(\alpha)}$,
\begin{eqnarray}
\bb{v}^{(\alpha)}&=(1,-1,1,-1,\ldots)^{T}=\sum_{k=0}^{\alpha}(-1)^k\ket{k},\\
\bb{u}^{(\alpha)}&=(0,-1,2,-3,\ldots,(-1)^{\alpha}\alpha)^{T}=\sum_{k=0}^{\alpha}(-1)^{k}k\ket{k},
\label{nullvectors}
\end{eqnarray}
such that
\begin{equation}
\bb{H}^{(\alpha)}\bb{v}^{(\alpha)}=0,\qquad (\bb{H}^{(\alpha)})^{2}\bb{u}^{(\alpha)}=0,\qquad \bb{H}^{(\alpha)}\bb{u}^{(\alpha)}=
\frac{\alpha}{x}\bb{v}^{(\alpha)}
\end{equation}
In order to prove our case (\ref{exampleD2},\ref{nullspacevector}), we require another set of identities, expressing the action of a squared generator
$\alpha$-blocks $(\bb{H}^{(\alpha)})^{2}$ on the vector $\bb{u}^{(\alpha)}$, and transformation of both $\bb{u}^{(\alpha)}$ and $\bb{v}^{(\alpha)}$ under the
action of $\bLambda^{(\alpha)s}_{1,2}$. For the sake of brevity, we entirely omit their justification here (it can be found in \ref{KerApp}):
\begin{eqnarray}
\bLambda^{(\alpha)0}_{1}\bb{v}^{(\alpha)}&=\bLambda^{(\alpha)+}_{2}\bb{v}^{(\alpha+1)}=0,\nonumber \\
\bLambda^{(\alpha)\z}_{1}\bb{v}^{(\alpha)}&=-2\alpha\bb{v}^{(\alpha)},\nonumber \\
\bLambda^{(\alpha)+}_{1}\bb{v}^{(\alpha+1)}&=\alpha \bb{v}^{(\alpha)},
\label{auxv}
\end{eqnarray}
and
\begin{equation}
\bLambda^{(\alpha)+}_{2}\bb{u}^{(\alpha+1)}=2\bb{v}^{(\alpha)}.
\label{auxu}
\end{equation}
Accounting for expansion of $\bb{D}^{(\alpha)+}_{2}$ \eref{exampleD2}, and using auxiliary identities \eref{auxv},\eref{auxu},
we show $\bb{D}^{(\alpha)+}_{2}\bb{u}^{(\alpha+1)}=0$. Linear independence of $\bb{u}^{(\alpha+1)}$ from the column space of $\bLambda^{(\alpha)-}_{1}$ therefore allows
for its extension to an invertible matrix $\tilde{\bLambda}^{-}_{1,\alpha}$ by adding $\bb{u}^{(\alpha+1)}$ as the $(\alpha+2)$-th column vector, yielding
\begin{equation}
\bb{D}^{(\alpha)+}_{2}=\bLambda^{(\alpha-1)-}_{1}\bb{D}^{(\alpha-1)+}_{2}(\tilde{\bLambda}^{(\alpha)-}_{1})^{-1},
\end{equation}
from where it immediately follows $\bb{D}^{(\alpha)+}_{2}=0$ if $\bb{D}^{(\alpha-1)+}_{2}=0$.\\

An entirely analogous reasoning applies to the remaining four cases from \eref{Doperators}. Using null-vector of $(\bb{H}^{(\alpha)})^{2}$
we derive the action of the operators $\{\bLambda^{s}_{1,\alpha}\}$ for $s=\{0,+,z\}$,
\begin{eqnarray}
\bLambda^{(\alpha)0}_{1}\bb{u}^{(\alpha)}&=-2\alpha \bb{v}^{(\alpha)},\\
\bLambda^{(\alpha)\z}_{1}\bb{u}^{(\alpha)}&=-2\alpha \bb{v}^{(\alpha)}-2(\alpha-2)\bb{u}^{(\alpha)},
\end{eqnarray}
which justifies adding $\bb{u}^{(\alpha)}$ (or $\bb{u}^{(\alpha+1)}$ in the case of $\bb{D}^{(\alpha)+}_{1}$) to the columns of $\bLambda^{(\alpha)-}_{1}$ when
operating by the corresponding $\bb{D}^{(\alpha)s}_{1}$ from the left. Essentially it sufficient to demonstrate that $\bb{D}^{(\alpha)s}_{k}$ preserve the null-space
$\ker(\bb{H}^{2})$.

In the case of diagonal blocks $\bb{D}^{(\alpha)0}_{1}$ and $\bb{D}^{(\alpha)\z}_{1}$, after multiplying them by the conserved charge
$\bLambda^{(\alpha)-}_{1}$ from the right, we use identities \eref{lambdaidentities} and
\begin{equation}
\bb{H}^{(\alpha)}\bLambda^{(\alpha-1)-}_{1}=\bLambda^{(\alpha-1)-}_{1}\bb{H}^{(\alpha-1)}
\end{equation}
to bring $\bLambda^{(\alpha)-}_{1}$ to the left, finishing with
\begin{equation}
\bb{D}^{s}_{1,\alpha}\bLambda^{(\alpha)-}_{1}=\bLambda^{(\alpha-1)-}_{1}\bb{D}^{(\alpha-1)s}_{1},\qquad s\in\{0,\z\}.
\end{equation}
An extra linear term in $\bb{D}^{(\alpha)\z}_{1}$ is of no importance, as it cancels
out regardless of its prefactor. Finally we take care of $\bb{D}^{(\alpha)+}_{1}$, arriving at the following coupled operator recursion
\begin{equation}
\bb{D}^{(\alpha)+}_{1}\bLambda^{(\alpha)-}_{1}=\bLambda^{(\alpha-1)-}_{1}\bb{D}^{(\alpha)+}_{1}+\bb{D}^{(\alpha)\z}_{1}.
\label{eqn:coupled}
\end{equation}
We have nevertheless already proven that $\bb{D}^{(\alpha)\z}_{1}=0$ for every $\alpha$, hence the recurrence becomes of the same type as the ones above.

\section{Properities of the exterior integrability structures}

\subsection{Properties of the $R$-matrix}

$R$-matrices are required to obey additional compatibility-type condition (the celebrated \textit{Yang-Baxter equation}, or in our notation, the braid group relation)
imposed on a triple-product of auxiliary spaces ${\cal H}^{\otimes 3}_{\rm a}$,
\begin{equation}
\!\!\!\!\!\!\!\!\!\!\!\!\!\!\!\!\!\!\!\!\!\!\!\!\!\!\!\!\!\!\!(\mathds{1}\otimes \bb{R}(\lambda,\mu))(\bb{R}(\lambda,\eta)\otimes \mathds{1})(\mathds{1}\otimes \bb{R}(\mu,\eta))=
(\bb{R}(\mu,\eta)\otimes \mathds{1})(\mathds{1}\otimes \bb{R}(\lambda,\eta))(\bb{R}(\lambda,\mu)\otimes \mathds{1}),\;
\label{BraidYangBaxter}
\end{equation}
which automatically ensures associativity of intertwining property over multiple spaces ${\cal H}_{\rm a}$. Despite in this paper we only strictly
prove the $RLL$ relation (\ref{RLL}) we discuss in Section \ref{discussion} some other related results \cite{K01,K02} from which (\ref{BraidYangBaxter}) should also follow.

Unlike in most often encountered cases of integrable models (e.g. in fundamental models),
the $R$-matrix here is \textit{not of difference type}, i.e. its elements do not depend on the difference of the involved spectral parameters only.
Yet, the difference of spectral parameters, curiously enough, enters in a way (\ref{expform}) which is reminiscent of a Lie group structure.

Additionally, one observes the following interesting properties of the $R$ operator (all following directly from explicit representation
 $\bb{R}(\lambda,\mu)=\exp\left((\lambda-\mu)\bb{H}\left((\lambda+\mu)/2\right)\right)$ and the properties of the generator $\bb{H}(x)$).
 \begin{enumerate}
 \item {\em Regularity}:
 \begin{equation}
 \bb{R}(\lambda,\lambda)=\mathds{1}.
 \end{equation}
 \item  {\em  $P$-symmetry} [See definition (\ref{pia})]:
\begin{equation}
\pi_{\rm a}(\bb{R}(\lambda,\mu)) =\bb{R}(\mu,\lambda).
\end{equation}
\item {\em Orthogonality}:
\begin{equation}
\bb{R}(\lambda,\mu)\bb{R}(\mu,\lambda)=\one.
\end{equation}
\item From (ii) and (iii) we immediately derive another nice property, namely that the eigenvalues of the matrix $\mm{P}\mm{R}(\lambda,\mu)$ can only be $\pm 1$, namely
\begin{equation}
(\mm{P}\mm{R}(\lambda,\mu))^2=\one
\end{equation}
\item All eigenvalues of $\mm{R}(\lambda,\mu)$ are equal to $1$. The $R$ operator has therefore a non-trivial {\em Jordan canonical form}, where each $(\alpha+1)\times (\alpha+1)$ matrix
$\mm{R}^{(\alpha)}(\lambda,\mu)$ is similar to a single irreducible Jordan block [following from (\ref{jordan})].

\item The matrix $\bb{R}(x+\frac{1}{2}y,x-\frac{1}{2})$ is {\em holomorphic} in both $x,y\in\CC$ except at $x\in \frac{1}{2}\ZZ^+$ where it has {\em simple poles}.
This follows from the fact that the generator $\bb{H}(x)$ has simple poles at  $x\in \frac{1}{2}\ZZ^+$ as well, property (v) which terminates the exponential series after $\alpha+1$ terms in subspace
 ${\cal H}^{(\alpha)}_{\rm a}$, and a curious nilpotent algebra among its residua $\bb{X}^{(p)}={\rm Res}_{x=\frac{p}{2}}\bb{H}(x)$ \eref{Y},
namely
\begin{equation}
\mm{X}^{(p)} \mm{X}^{(m)} = 0 \quad {\rm if}\quad p\ge m.
\label{xx}
\end{equation}
The property (\ref{xx}) can be studied in each space ${\cal H}^{(\alpha)}_{\rm a}$ separately, where it can be proven by application of inductive arguments on elementary binomial identities.
\item
At the poles, actually, where $\lambda+\mu \in \ZZ^+$, the intertwining of the product of Lax operators can be implemented by taking a residuum of the $RLL$ relation (\ref{RLL}).
\item
Comparing the representation (\ref{MPOtensors}) of $\mathfrak{sl}(2)$ with the transposed one\footnote{Transposition is defined, as usual, $(\ket{k}\bra{l})^T\equiv \ket{l}\bra{k}$, without complex conjugation.}, which should be equivalent
\begin{equation}
(-1)^s \mm{A}^T_s(\lambda) = \mm{U}(\lambda)\mm{A}_{-s}(\lambda)\mm{U}^{-1}(\lambda),\quad s\in\{+,0,-\},
\label{AT}
\end{equation}
where $\mm{U}(\lambda)$ is  a {\em diagonal} operator from ${\rm End}({\cal H}_{\rm a})$ (invertible for $\lambda \not\in \frac{1}{2}\ZZ^+$),
\begin{equation}
U^k_l(\lambda) = \delta_{k,l} {2\lambda \choose k},
\end{equation}
we obtain the corresponding transposal symmetry for the $R$-matrix
\begin{equation}
(\mm{U}(\lambda)\otimes\mm{U}(\mu))\mm{R}(\lambda,\mu)(\mm{U}^{-1}(\mu)\otimes\mm{U}^{-1}(\lambda)) =  \mm{R}^T(\mu,\lambda).
\end{equation}
Note that this is a kind of Liouvillian $\mathbb{P}\mathbb{T}$ symmetry of the type proposed in Ref.~\cite{P12b}.
The sign factor $(-1)^s$ in (\ref{AT}) is a consequence of non-canonical (real) representation of (\ref{MPOamplitudes}).
\end{enumerate}

\subsection{Properties of the monodromy matrix}

Rich structure and properties of the $R$ operator discussed above are also inherited by the corresponding exterior monodromy operator $\bb{T}(\lambda)$ or its matrix elements (\ref{meT}).
We list some of the most remarkable properties that we have observed here, with the hope that they will find useful future applications (e.g. those discussed in Section \ref{discussion}).

Firstly, for a given system size $n$,  selection rule (\ref{selrule}) implies that the monodromy matrix is {\em banded}, i.e. 
\begin{equation}
T^k_l(\lambda)=0,\quad {\rm if} \quad |k-l| > n.
\end{equation}
Furthermore, we claim that only the elements $T^k_l(\lambda)$ from a $(n+1)\times (n+1)$ square, namely for $k\le n$, $l\le n$ are {\em linearly dependent} physical operators.
Therefore, for a fixed distance from the diagonal $q=|k-l|$, only $n-q$ matrix elements are linearly independent, while all others can be expressed in terms of those
\begin{equation}
T^l_{l+q}(\lambda) = \sum_{k=0}^{n-q} c^+_{n,q,l,k}(\lambda)T^k_{k+q}(\lambda), \;\; T^{l+q}_{l}(\lambda) = \sum_{k=0}^{n-q} c^-_{n,q,l,k}(\lambda)T^{k+q}_{k}(\lambda),
\end{equation}
where $c^\pm_{n,q,l,k}(\lambda)$ are some rational functions of $\lambda$ with integer coefficients.

Secondly, we were looking for linear combinations of magnetization (particle-number) preserving diagonal matrix elements $T^k_k(\lambda)$ that would form a commuting family.
Up to linear dependences we conjecture (based on empirical evidence) that there exists a single commuting linear combination besides $S(\lambda)=T^0_0(\lambda)$, namely
\begin{equation}
\tilde{S}(\lambda) = \sum_{k=1}^{n} (-1)^{n+k}{n \choose k} \frac{2\lambda-n+1}{2\lambda-k+1} T^k_k(\lambda) ,
\end{equation}
so that
\begin{equation}
[\tilde{S}(\lambda),\tilde{S}(\mu)]=0,\quad [\tilde{S}(\lambda),S(\mu)]=0.
\end{equation}

Thirdly, as in our problem we are dealing with non-normal operators one may want to understand the connection between the transposed ($[T^l_k(\lambda)]^T$)
and original ($T^k_l(\lambda)$) monodromy elements. Writing the {\em reflection} parity operator in quantum spin space $Q=Q^{-1}\in {\rm End}({\cal H}^{\otimes n}_{\rm s})$,
$Q\ket{\nu_1,\nu_2,\ldots,\nu_n} =\ket{\nu_n,\nu_{n-1},\ldots,\nu_1}$, we find immediately [applying  Eqs. (\ref{Tmatrix},\ref{AT})]
\begin{equation}
\left(T^l_k(\lambda)\right)^T = (-1)^{k-l} {2\lambda \choose k}{2\lambda \choose l}^{\!-1} Q T^k_l(\lambda) Q,
\end{equation}
or more compactly\footnote{Yet, another form of $\mathbb{P}\mathbb{T}$-like symmetry \cite{P12b}.},
writing a {\em partial transpose} with respect to ${\cal H}^{\otimes n}_{\rm s}$ by superscript $T_{\rm s}$,
\begin{equation}
\mm{T}^{T_{\rm s}}(\lambda) = \tilde{\mm{U}}(\lambda)Q \bb{T}(\lambda)Q \tilde{\mm{U}}^{-1}(\lambda),
\end{equation}
were $\tilde{\mm{U}}(\lambda) = {\rm diag}(1,-1,1,-1\ldots)\mm{U}(\lambda)$.
Furthermore, the reflected monodromy elements $Q T^{k}_l(\lambda) Q$ can be in turn expressed in terms of linear combination of $T^{j}_{l-k+j}(\lambda)$.
For example, we state the connection explicitly for $00$ matrix element
\begin{equation}
\left(T^0_0(\lambda)\right)^T = {2\lambda \choose n+1}\sum_{k=0}^n {n+1\choose k+1}\frac{(-1)^k (k+1)}{2\lambda - k} T^k_k(\lambda).
\end{equation}

Finally, let us consider the action of monodromy elements on spin states with a small number of quasi-particle excitations.
Let $\ket{\Omega_m}\in{\cal H}^{\otimes n}_{\rm s}$ denote an arbitrary state from ${n \choose m}$ dimensional subspace with {\em exactly} $m$ spins-up (and all other $n-m$ spins down), i.e.
$M\ket{\Omega_m} = (2m-n)\ket{\Omega_m}$, and let $\ket{\tilde{\Omega}_m}$ denote a state with  $m$ down-spins, $\ket{\tilde{\Omega}_m}\equiv \ket{\Omega_{n-m}}$.
Direct inspection using explicit representation (\ref{MPOamplitudes}) reveals the action on the vacuum state in terms of a {\em shift} of spectral parameter
\begin{eqnarray}
T^l_{l+q}(\lambda)\ket{\Omega_0}&=&{2\lambda-l\choose q}{2\lambda-2l\choose q}^{\!-1} T^0_q(\lambda-k)\ket{\Omega_0}, \\
T^{l+q}_l(\lambda)\ket{\tilde{\Omega}_0}&=&{l+q\choose l} T^q_0(\lambda-k)\ket{\tilde{\Omega}_0}.
\end{eqnarray}
whereas, one can write similar but more general expressions for the $m$-particle sectors
\begin{eqnarray}
T^l_{l+q}(x) \ket{\Omega_{m}} &=& \sum_{k=0}^{m} r^{q,m}_{l,k}(\lambda) T^k_{k+q}\left(\lambda - (l-k))\right)\ket{\Omega_{m}}, \\
T^{l+q}_l(x)\ket{\tilde{\Omega}_{m}} &=& \sum_{k=0}^{m} s^{q,m}_{l,k}(\lambda) T^k_{k+q}\left(\lambda - (l-k)\right)\ket{\tilde{\Omega}_{m}},
\end{eqnarray}
where $q\in\ZZ^+$ can be interpreted as the number of quasi-particles created and $l>m$ for the relations to be non-trivial.
Remarkably, the rational functions  $r^{q,m}_{l,k}(\lambda)$, $s^{q,m}_{l,k}(\lambda)$, again having integer coefficients, and only simple poles at $\frac{1}{2}\ZZ^+$, do not depend on system size $n$.
For the purpose of treating the NESS density operator, say for developing an ABA procedure for diagonalizing it, it should be handy to control
transposed matrix elements at negative spectral parameter $-\lambda$, corresponding to Hermitian conjugation at real value of dissipation $\varepsilon=2\ii/\lambda$, for which $\bar{\lambda}=-\lambda$ and $S^\dagger(\lambda)=S^T(-\lambda)$. Let us write
\begin{equation}
\tilde{T}^l_k(x) := (-1)^n [T^k_l(-x)]^T,
\end{equation}
where a sign factor $(-1)^n$ is put for convenience.
Then, straightforward inspection again suggests remarkable connections:
\begin{eqnarray}
T^l_{l+q}(\lambda) \ket{\Omega_{m}} &=& \sum_{k=0}^{m} f^{q,m}_{l,k}(\lambda) \tilde{T}^{k+q}_k\left(\lambda - (q+l+k)\right)\ket{\Omega_{m}},\label{conn1} \\
\tilde{T}^{l+q}_l(\lambda)\ket{\Omega_{m}} &=& \sum_{k=0}^{m} g^{q,m}_{l,k}(\lambda) T^k_{k+q}\left(\lambda + (q+l+k)\right)\ket{\Omega_{m}}, \label{conn2}\\
T^{l+q}_{q}(\lambda) \ket{\tilde{\Omega}_{m}} &=& \sum_{k=0}^{m} g^{q,m}_{l,k}(-\lambda) \tilde{T}^{k}_{k+q}\left(\lambda - (q+l+k)\right)\ket{\tilde{\Omega}_{m}}, \\
\tilde{T}^{l}_{l+q}(\lambda)\ket{\tilde{\Omega}_{m}} &=& \sum_{k=0}^{m} f^{q,m}_{l,k}(-\lambda) T^{k+q}_{k}\left(\lambda + (q+l+k)\right)\ket{\tilde{\Omega}_{m}},
\end{eqnarray}
where relations are already non-trivial for any $l$, and integer coefficient rational functions $f^{q,m}_{l,k}(\lambda)$,  $g^{q,m}_{l,k}(\lambda)$ again do not depend on size $n$.

\section{Discussion}

\label{discussion}

After this work has been completed, we have learned about Refs.~\cite{K01,K02,K12} where related infinitely-dimensional $R$-matrices have been constructed using manifestly $\mathfrak{sl}(2)$-symmetric form of Lax and $R$-matrices. It seems that such a universal  $\mathfrak{sl}(2)$ $R$-matrix might be useful in the context of QCD and high-energy physics whereas in condensed matter physics the non-unitarity of the general infinitely-dimensional representation seems to be only compatible with phenomena far from equlibrium which we discuss here.

In fact, our Lax matrix  (\ref{Lax}) becomes  $\mathfrak{sl}(2)$-symmetric after multiplying by $\sigma^\z$, $\tilde{\mm{L}}(\lambda) = \mm{L}(\lambda)\sigma^\z = \mm{B}_+(\lambda)\otimes \sigma^- +  \mm{B}_-(\lambda)\otimes \sigma^+ + \mm{B}_\z(\lambda)\otimes \sigma^\z = \vec{\mm{B}}(\lambda)\cdot \vec{\sigma}$,
where $\mm{B}_+ = \mm{A}_-, \mm{B}_-=-\mm{A}_+, \mm{B}_\z = \mm{A}_0$ are {\em canonical} generators of infinitely dimensional representation of $\mathfrak{sl}(2)$ with representation parameter $\lambda$ \footnote{In the usual complex representation we in addition have to re-define the generators $\mm{B}_\pm \to -\ii \mm{B}_\pm$.}.
The $R$-matrices resulting from solving $RLL$ relations for the two forms of $L$-matrices, $\mm{L}(\lambda)$ and $\tilde{\mm{L}}(\lambda)$, are different but closely related. Nevertheless, the results presented in this paper are more explicit and detailed in connection to a different form of a transfer matrix as they are taylored for  non-equilibrium condensed matter applications, and hence they are essentially non-overlapping with those of Ref.~\cite{K01}. Although the $\mathfrak{sl}(2)$-symmetric $L$-matrix generates a related transfer matrix, namely $S(\lambda)(\sigma^\z)^{\otimes n}$, and yields an identical NESS density operator
$S(\lambda)S^\dagger(\lambda)$, we have a good reason to use also a symmetry-broken representation of the Lax matrix. Namely, only in our representation the MPA for $S(\lambda)$ generates a convergent sum of {\em local operators} \cite{IP13} in the $q$-deformed case of anisotropic $XXZ$ model (see discussion below, in subsect. \ref{xxzsection}).

We foresee two immediate interesting applications of the exterior (non-equilibrium) integrability formulated here.

\subsection{Algebraic Bethe Ansatz and spectrum of the density operator}

A tempting proposal following from our construction is the construction of ABA procedure for diagonalizing NESS density operator. This could be particularly interesting in the light of recent suggestions \cite{MPCP13,PZ13} that the spectral properties of  equilibrium and non-equilibrium density operators can be used as indicators of integrability (or exact solvability) similar as in the idea of {\em quantum chaos}.

Algebraic form of Bethe ansatz allows for construction of an eigensystem for a family of mutually commuting transfer operators. The procedure is
based on the quasi-particle modes created under the action of (off-diagonal) elements of the monodromy matrix $\bb{T}(\lambda)=\bb{L}(\lambda)^{\otimes_{\rm s} n}$.
Many-particle excitations arise as a string of monodromy elements, operating on a specially chosen reference state. The role of
$R$-matrix is to prescribe quadratic algebraic relations among elements with different value of spectral parameter (which are interpreted
as quasiparticle momenta), enabling for construction of eigenstates of the quantum transfer operator. A set of $n$ spectral parameters
$\{\lambda_k\}$ for $n$-particle excitations has to be chosen accordingly in order to eliminate unwanted terms (those that are not the eigenvectors)
which unavoidably emerge during commutation of the elements of $\bb{T}$. The latter condition gives rise to famous Bethe ansatz equations \cite{B31}.

As construction of ABA in this case, due to Cholesky structure of the diagonalizing operator (\ref{Choleskyform}), does not seem to be straightforward, we outline here only the first step. Namely on how to obtain single quasi-particle excitations,
i.e. eigenvalues and eigenvectors of $\tilde{\rho}_\infty(\lambda)=S(\lambda) S^T(-\lambda)= (-1)^n T^{0}_{0}(\lambda)\tilde{T}^{0}_{0}(\lambda)$, where $\lambda = 2\ii/\varepsilon \in \ii\RaR$ of the type $\ket{\Omega_1}$.
Applying the connections (\ref{conn1},\ref{conn2}) and the $RTT$ relation (\ref{RTT}) for ${\cal H}^{(\alpha=1)}_{\rm a}$ sector only, therefore using only the $2\times 2$ block $\mm{R}^{(1)}$, we arrive at the useful identity
\begin{eqnarray}
& (-1)^n \tilde{\rho}_\infty(\lambda)T^0_1(\mu) \ket{\Omega_0} = [t(\lambda)]^2 \Lambda(\lambda,\mu)\, T^0_1(\mu) \ket{\Omega_0} \label{UWT}\\
&+ \frac{\mu((\lambda+\mu-1)t(\lambda)t(\mu) - 2\lambda(\mu-\lambda) t(\lambda+1)t(\mu-1))}{(\mu-\lambda)(\lambda-\mu+1)}\, T^0_1(\lambda) \ket{\Omega_0}    \nonumber \\
&+ \frac{2\mu\lambda(\lambda+1/2) t(\lambda) t(\mu-1)}{(\lambda+1)(\lambda-\mu+1)}\,  T^0_1(\lambda+1) \ket{\Omega_0},    \nonumber
\end{eqnarray}
where $t(\lambda):=\lambda^n$ and $\Lambda(\mu,\lambda)$ is a quasi-particle dispersion relation:
\begin{equation}
\Lambda(\mu,\lambda) := \frac{(\lambda+\mu)(\lambda+\mu-1)}{(\lambda-\mu)(\lambda-\mu+1)}.
\end{equation}
There are two single quasi-particle states with the same eigenvalue $\Lambda(\mu_1,\lambda)=\Lambda(\mu_2,\lambda)$, which can be parametrized in terms of a single complex {\em rapidity} parameter $\xi$, as $\mu_1=\frac{1}{2}(1 + (\lambda+1) \xi)$, $\mu_2=\frac{1}{2}(1 + (\lambda-1)/\xi)$.
Hence the single-particle ABA is already a nontrivial combination of two terms
\begin{equation}
\ket{\Psi} = (C_1 T^0_1(\mu_1) + C_2 T^0_1(\mu_2))\ket{\Omega_0}
\end{equation}
As the three operators on the RHS of (\ref{UWT}) are {\em linearly independent}, the requirement that the two {\em unwanted terms}, proportional to vectors
$T^0_1(\lambda) \ket{\Omega_0}$ and $T^0_1(\lambda+1) \ket{\Omega_0}$, cancel, i.e. to have
$\rho_{\infty}(\lambda)\ket{\Psi} = \Lambda \ket{\Psi}$, results in requiring that a $2\times 2$ system of equations for $C_1,C_2$ has a nontrivial solution, i.e.
\begin{equation}
\left( \frac{1 - (\lambda+1)\xi}{1+(\lambda+1)\xi}\right)^n  \left(\frac{\xi + \lambda-1}{\xi - \lambda+1}\right)^n = \left(\frac{1-\xi}{1+\xi}\right)\frac{(\lambda+1)\xi + \lambda-1}{(\lambda+1)\xi - \lambda+1}.
\end{equation}
This can be understood as a Bethe equation for single particle eigenvectors of NESS, with eigenvalue $\Lambda(\frac{1}{2}(1+(\lambda+1)\xi),\lambda)$.

However, generalizing this procedure to multiple excitations seems far from trivial and should be a challenge for future work.

\subsection{The anisotropic $XXZ$ model and a new family of quasi-local conservation laws}
\label{xxzsection}

As has been pointed out in Ref.~\cite{K02}, the infinitely dimensional $R$-matrix also exists for continuous representations of the quantum group $U_q(\mathfrak{sl}(2))$
hence all our constructions of exterior integrability should be $q-$deformable and should translate to the boundary-driven anisotropic $XXZ$ spin chain \cite{P11a,P11b,KPS13} where the Hamiltonian density $h$ in (\ref{xxx}) should be replaced by $h=2\sigma^+\otimes \sigma^{-}
+2\sigma^{-}\otimes \sigma^{+} + \Delta \sigma^{\z}\otimes \sigma^{\z}$ with the anisotropy parameter $\Delta$.

Most interesting there is the question, whether the recently discovered quasi-local conservation law \cite{P11a} can be generalized and extended to a whole family.
In integrable theories, local conserved quantities are usually obtained in terms of logarithmic derivatives of transfer matrices around some trivial values of the spectral parameter. Here, the spectral parameter is non-standard and is related to coupling to the environment, hence the derived conserved quantities can have different spin-flip symmetry $K = (\sigma^\x)^{\otimes n}$ as in the standard case
\cite{F94,KBI93} where due to the equivalence of quantum spin and auxilliary spaces the $K$ symmetry is imposed to the $L$- and $R$- matrices as well and henceforth to all so-derived families of conservation laws. In the exterior integrability problem, however, the $K$-symmetry is explicitly broken, resulting in (potentially quasi-local) conserved quantities which may yield non-trivial  Drude-weight bounds \cite{IP13} even in the absence of external magnetic field.

For example, writing for the moment the commuting transfer matrix (\ref{MPA},\ref{commut}) as a function
of dissipation $\varepsilon$ (in notation of Ref.~\cite{P11b}), which is polynomial  for finite $n$, the conservation law $Z$ which has been proposed and implemented in Ref.~\cite{P11a}
and which is quasi-local for $|\Delta|<1$ is simply $Z = \frac{\dd}{\dd \varepsilon} S(\varepsilon)\vert_{\varepsilon=0}$.
We conjecture that a further tower of (quasi-local, in case $|\Delta|<1$) conservation laws which break the $K$-symmetry is given by higher logarithmic derivatives
\begin{equation}
Z_{k} = \frac{\dd^{2k-1}}{\dd \varepsilon^{2k-1}} \log S(\varepsilon)\vert_{\varepsilon=0},\quad k=1,2,\ldots
\end{equation}
The even order logarithmic derivatives vanish as a consequence of an interesting identity
\begin{equation}
S^{-1}(\varepsilon)=S(-\varepsilon),
\end{equation}
which can be easily proven. Details on these constructions shall be presented elsewhere.

\subsection{Conclusion}

We have provided a new link between the matrix product ansatz and Yang-Baxter integrability in the context of non-equilibrium quantum physics, which is fundamentally different than the one which exists on the level of closed quantum systems \cite{AL04}. 
The first fundamental difference is in the role of spectral parameter of the integrable theory which is now taken by a continuous representation parameter of infinite-dimensional representation of the underlying quantum symmetry of the model. The second fundamental difference is the formulation of the transfer matrix, which is here, due to infinite-dimensionality of the representation space, taken by the ground-state expectation instead of a trace.
Generalizations to other quantum integrable models seem straightforward, the most obvious one being perhaps the multi-component quantum hopping model \cite{S75}.

\section*{Acknowledgements}

We thank Marko Petkov\v sek for useful discussions  on computer algorithms for proving identities involving binomial symbols.
E. I. thanks David Gajser for presenting the proof for the property (\ref{xx}).
The work was supported by the Grant P1-0044 of Slovenian Research Agency.

\appendix
\section{Explicit expression of the generator $H(x)$}

\label{appH}
Here we provide three additional, trivially equivalent, but useful forms for the $\alpha$-block of the generator $\bb{H}^{(\alpha)}(x)$ of the exterior $R$-matrix.

\paragraph*{Compact form.}
\begin{eqnarray}
H^{(\alpha)}_{k,l}(x)=\frac{(-1)^{k-l}}{k-l}{k \choose l}{k-1-2x \choose k-l}^{-1},\qquad k\geq l+1,\\
H^{(\alpha)}_{k,k}(x)=-\frac{1}{2}\frac{d}{dx}\log{\alpha-k-1-2x \choose \alpha-2k},\qquad 2k\leq \alpha,
\label{secondform}
\end{eqnarray}
and \eref{firstform} elsewhere, where use the $\mathbb{C}$-number extension of the binomial symbol
\begin{equation}
{x \choose k}:=\frac{x(x-1)(x-2)\cdots (x-k+1)}{k(k-1)(k-2)\cdots 1},\qquad x\in \mathbb{C}.
\label{extendedbinomial}
\end{equation}
Showing that (\ref{secondform}) is identical to the expressions (\ref{firstform}) amounts to checking that they have identical residua.
This in turn leads us to the third,

\paragraph*{Residue form.}
For most purposes we find useful the following matrix-valued residue decomposition of $\bb{H}^{(\alpha)}$,
with poles positioned at $x=m/2$, $m=0,1,2,\ldots$:
\begin{eqnarray}
\bb{H}^{(\alpha)}(x)&=\sum_{m=0}^{\alpha}\bb{X}^{(\alpha)m}f_{m}(x),\\
\bb{X}^{(\alpha)m}&:={\rm Res}_{x=\frac{m}{2}}\bb{H}^{(\alpha)}(x),
\label{residuedecomposition}
\end{eqnarray}
and matrix coefficients ${X^{(\alpha),m}_{k,l}}$ are given in $P$-symmetric form explicitly as [see (\ref{firstform})]
\begin{eqnarray}
X^{(\alpha)m}_{k,l}&=\frac{1}{2}\left(Y^{(\alpha)m}_{k,l}-Y^{(\alpha)m}_{\alpha-k,\alpha-l}\right),\\
Y^{(\alpha)m}_{k,l}&=(-1)^{k-m-1}{k \choose l}{k-l-1 \choose m-l} \theta_{m-l}. \label{Y}
\label{residueform}
\end{eqnarray}
where $\theta_x :=1$ if $x\ge 0$ and $\theta_x:=0$ if $x < 0$.
Note that in terms of parity operator (\ref{P}):
\begin{equation}
\bb{X} = \frac{1}{2}(\bb{Y} - \bb{P}\bb{Y}\bb{P}).
\label{XP}
\end{equation}

\paragraph*{Jordan form.}
Here is the form which is in fact equivalent to a Jordan decomposition of $\bb{H}^{(\alpha)}$.
Let $\mm{W}^\alpha(x)$ be an {\em upper triangular} $(\alpha+1)\times(\alpha+1)$ matrix with entries:
\begin{equation}
W^\alpha_{k,l}(x) = (-1)^{k+l} 2^{l-\alpha}{\binom{\alpha}{l}}^{-1}\binom{\alpha-k}{\alpha-l}\binom{2x}{\alpha-l}
\end{equation}
which vanish if $k > l$,
and $\mm{\Delta}^\alpha$ a {\em strictly lower triangular} matrix with constant entries:
\begin{eqnarray}
\Delta^\alpha_{k,l} &=& \frac{2^{l-k+1}}{k-l}   \quad {\rm if} \qquad k > l, \\
\Delta^\alpha_{k,l} &=& 0 \,\quad\qquad{\rm if} \qquad k \le l.
\end{eqnarray}
Then, we have the following decomposition:
\begin{equation}
\mm{H}^\alpha(x) = \mm{W}^\alpha(x) \mm{\Delta}^\alpha \left(\mm{W}^\alpha(x)\right)^{-1},
\label{jordan}
\end{equation}
where, noting, $\left(\mm{W}^\alpha(x)\right)^{-1}$ is again upper triangular.

\section{Verification of the HLL relation.}
\label{appHLL}

By utilizing the residue decomposition \eref{residuedecomposition} of $\alpha$-block $\bb{H}^{(\alpha)}(x)$ in terms of $\bb{X}^{(\alpha)}$,
we calculate component-wise expansions of
\begin{equation}
[\bb{H}^{(\alpha)}(x),\bLambda_{0}(x)]=\bLambda_{1}
\label{HLLappendix}
\end{equation}
using definition \eref{lambdas} with explicit form of MPA amplitudes \eref{MPOamplitudes}. Let us initially consider the diagonal
physical components $E^{00}$ (recall that $E^{11}$ is automatically obeyed by virtue of symmetry (\ref{HpiH2})), where block operators
$\bLambda^{(\alpha)00}_{0}$ preserve ${\cal H}^{(\alpha)}_{\rm a}$. It is also noteworthy that the corresponding $\bLambda^{(\alpha)00}_{1}$ on the right side is
bidiagonal (constant) matrix, therefore a direct calculation leads to, after isolating matrix coefficients in front of every simple pole $f_{p}(x)$,
$p=0,1,\ldots, \alpha$ and for every matrix element $k,l=0,1,\ldots,\alpha$
\begin{eqnarray}
(k-l)(k+l-\alpha)X^{(\alpha)p}_{k,l}&+(\alpha-l+1)(l-1-p)X^{(\alpha)p}_{k,l-1}\nonumber \\
&+(k-\alpha)(k-p)X^{(\alpha)p}_{k+1,l}=0.
\label{diagonalpoles}
\end{eqnarray}

When off-diagonal physical components are considered, two adjacent sectors $\alpha$ and $\alpha+1$ will get coupled,
i.e. for arbitrary $\alpha$ we evaluate
\begin{equation*}
[\bb{H}^{(\alpha)}(x),\bLambda^{(\alpha)01}_{0}(x)]=\bLambda^{(\alpha)01}_{1},\qquad
[\bb{H}^{(\alpha)}(x),\bLambda^{(\alpha)10}_{0}]=\bLambda^{(\alpha)10}_{1},
\end{equation*}
or in component notation,
\begin{eqnarray}
(l-p-1)(2l+p-2\alpha-2)X^{(\alpha)p}_{k,l-1}+(2l-p)(l+p-\alpha)X^{(\alpha)p}_{k,l}\nonumber \\
-(2k-p)(k+p-\alpha)X^{(\alpha+1),p}_{k,l}-(k-p)(2k+p-2\alpha)X^{(\alpha+1),p}_{k+1,l}=0,\\
k(2k+p-2\alpha-2)X^{(\alpha)p}_{k-1,l}+(2k-p)(k-\alpha-1)X^{(\alpha)p}_{k,l}\nonumber \\
-(2l-p)(l-\alpha-1)X^{(\alpha+1),p}_{k,l}-(l+1)(2l+p-2\alpha)X^{(\alpha+1),p}_{k,l+1}=0,
\label{E01poles}
\end{eqnarray}
for physical components $E^{01}$ and $E^{10}$, in respective order. All the relations above are of \textit{homogeneous} kind because the
singularities $\{f_{p}(x)\}$ are absent in the right-hand side of the $RLL$ relation (involving $\bLambda_{1}$). It is worth noticing that
parameter $p$ enters in the amplitudes from `fusion' with poles $f_{p}(x)$, by virtue of partial fraction expansion,
\begin{equation}
\frac{w(\alpha,k,l)-2x}{x-p/2}=\frac{w(\alpha,k,l)-p}{x-p/2}-2.
\end{equation}
For instance, working out \eref{diagonalpoles} explicitly
at fixed $\alpha$ and for the pole $f_{p}(x)$ (for sake of clarity we omit $x$-dependence from the amplitudes), we have
\begin{eqnarray*}
\fl \left([X^{(\alpha)p}f_{p},\bLambda^{00}_{0}(x)]\right)_{k,l}&=f_{p}\Big[
(a^{0}_{l}a^{0}_{\alpha-l}-a^{0}_{\alpha-k}a^{0}_{k})X^{(\alpha)p}_{k,l}
+a^{+}_{l-1}a^{-}_{\alpha-l}X^{(\alpha)p}_{k,l-1}-
a^{-}_{\alpha-k-1}a^{+}_{k}X^{(\alpha)p}_{k+1,l}\Big]\nonumber \\
&=f_{p}(k-l)(k+l-\alpha)X^{(\alpha)p}_{k,l}+f_{p}(l-1-p)(\alpha-l+1)X^{(\alpha)p}_{k,l-1}\nonumber \\
&+f_{p}(k-\alpha)(k-p)X^{(\alpha)p}_{k+1,l}\nonumber \\
&-2[(\alpha-l-1)X^{(\alpha)p}_{k,l-1}+(k-\alpha)X^{(\alpha)p}_{k+1,l}],
\end{eqnarray*}
Beside matrix-residue part, non-singular terms are produced as well. By collecting together non-singular contributions from all $\{f_{p}(x)\}$ and
matching them to non-vanishing (linear in $x$) terms on the right, we get an additional set of conditions which are to be satisfied:
\begin{equation}
\sum_{p=0}^{\alpha-1}(\alpha-l+1)X^{(\alpha)p}_{k,l-1}+(k-\alpha)X^{(\alpha)p}_{k+1,l}=\frac{1}{2}(2l-\alpha)\delta_{k,l}+
(l-\alpha-1)\delta_{k+1,l},
\end{equation}
at the $E^{00}$ component, and analogously
\begin{equation}
\sum_{p=0}^{\alpha}\left\{X^{(\alpha)p}_{k,l-1}+X^{(\alpha)p}_{k,l}-X^{(\alpha+1)p}_{k,l}-X^{(\alpha+1)p}_{k+1,l}\right\}=0,
\label{E01constant}
\end{equation}
\begin{eqnarray}
\sum_{p=0}^{\alpha}&\Big\{(2\alpha-l-p+1)X^{(\alpha)p}_{k,l-1}+(l-p+\alpha)X^{(\alpha)p}_{k,l}+(p-k-\alpha)X^{(\alpha+1)p}_{k,l}\nonumber \\
&+(k+p-2\alpha)X^{(\alpha+1)p}_{k+1,l}\Big\}=(3k-\alpha)\delta_{k,l}+(3k-2\alpha)\delta_{k+1,l},
\label{E01linear}
\end{eqnarray}
\begin{eqnarray}
\sum_{p=0}^{\alpha}&\Big\{kX^{(\alpha)p}_{k-1,l}-(k-\alpha-1)X^{(\alpha)p}_{k,l}+(l-\alpha-1)X^{(\alpha+1)p}_{k,l}\
-(l+1)X^{(\alpha+1)p}_{k,l+1}\Big\}\nonumber \\
&=(\alpha-l+1)\delta_{k,l}-(l+1)\delta_{k,l+1},
\label{E10constant}
\end{eqnarray}
at $E^{01}$ (equations \eref{E01constant},\eref{E01linear}) and $E^{10}$ (equation \eref{E10constant}).
However, as the latter set of expressions is rather tedious for further analytical manipulations, we decide at this point to take a different (however equivalent) strategy and rather employ the first form of the 
generator \eref{firstform}.

For the sake of compactness, we shall only provide explicit calculation to justify validity for the set of equations pertaining to non-singular part for physical components
$E^{00}$ and $E^{01}$, whereas an entirely equivalent procedure applies to show the identity associated with $E^{10}$ component.
We start with the diagonal element, where from \eref{HLLappendix} we obtain
\begin{eqnarray}
\label{HLL00expansion}
\left(\left[\bb{H}^{(\alpha)}(x),\bLambda^{(\alpha)00}_{1}\right]\right)_{k,l}&=(k-l)(k+l-\alpha)H^{(\alpha)}_{k,l}(x)\nonumber \\
&+(\alpha-l+1)(l-1-2x)H^{(\alpha)}_{k,l-1}\\
&+(k-\alpha)(k-2x)H^{(\alpha)}_{k+1,l}\nonumber \\
&=(\alpha-2l)\delta_{k,l}+2(\alpha-l+1)\delta_{k+1,l}=\left(\bLambda^{(\alpha)00}_{1}\right)_{k,l}.\nonumber
\end{eqnarray}
We introduce diagonal index $\delta:=k-l$ and focus initially on situation $\delta \geq 1$, where equations become homogeneous. Projecting
out components coupled to any $f_{p}(x)$ (as they are irrelevant for this part) we find the requirement
\begin{eqnarray}
\sum_{m=0}^{\delta}\Big[&(-1)^{m}{l+\delta \choose l-1}{\delta \choose m}(\alpha-l+1)\nonumber \\
&+(-1)^{m}{l+\delta+1 \choose l}{\delta \choose m}(l+\delta-\alpha)\Big]=0,\quad \delta\geq 1,
\end{eqnarray}
which is obviously true for all $l=0,1,\ldots,\alpha$, based on a well-known binomial identity
\begin{equation}
\sum_{m=0}^{\delta}(-1)^{m}{\delta \choose m}=0,\qquad \delta>0
\label{alternatingsum}
\end{equation}
The diagonal cases follow after plugging $\delta=0$ (beware of all the corresponding prefactors), where only $m=0$ contributes, yielding
\begin{equation}
-{l\choose l-1}(\alpha-l+1)-{l+1 \choose l}(l-\alpha)=\alpha-2l,
\end{equation}
which correctly reproduces diagonal elements of the right-hand side of \eref{HLL00expansion}. The same argument of course applies when
$\delta:=l-k\geq 2$, the only difference being the indices in the poles get reversed, i.e. the amplitudes fuse with the element
$H^{(\alpha)}_{\alpha-k,\alpha-l}$. This however leads to the same argument based on the identity \eref{alternatingsum} as long as the
non-singular part of the expression is considered only. Thus it remains to be checked in the case when $k=l-1$, where
\begin{eqnarray}
&(2l\!-\!\alpha\!-\!1)H^{(\alpha)}_{\alpha-l+1,\alpha-l}(x)+(\alpha\!-\!l\!+\!1)(l\!-\!1\!-\!2x)\left[H^{(\alpha)}_{l-1,l-1}(x)-H^{(\alpha)}_{l,l}(x)\right]\nonumber\\
&=\frac{1}{2}(2l-\alpha-1)(\alpha-l+1)f_{\alpha-l}-\frac{1}{2}(\alpha-l+1)(l-1-2x)(f_{l-1}+f_{\alpha-l})\nonumber
\\
&=2(\alpha-l+1),
\end{eqnarray}
where again the correct result $2(\alpha-l-1)\delta_{k+1,l}$ is reproduced.

The same procedure applies for the off-diagonal physical component, where element from two neighboring $\alpha$-subspaces are involved -- e.g. for
the $E^{01}$ we have to show that
\begin{eqnarray}
&(l-1-\alpha+x)(l-1-2x)H^{(\alpha)}_{k,l-1}(x)+(\alpha-l-2x)(-l+x)H^{(\alpha)}_{k,l}(x)\nonumber \\
&+(\alpha-k-2x)(k-x)H^{(\alpha+1)}_{k,l}(x)+(\alpha-k-x)(k-2x)H^{(\alpha+1)}_{k+1,l}\nonumber \\
&=(3k-\alpha)\delta_{k,l}+(3k-2\alpha)\delta_{k+1,l}.
\end{eqnarray}
By focusing once more on a non-singular part after resolving expansion in terms of $\{f_{m}(x)\}$ and terms which are now linear functions in
$x$, we find the vanishing of the latter is implied, for $\delta=k-l\geq 1$ and for $l-k\geq 2$ based according to \eref{alternatingsum} in
conjunction with another identity
\begin{equation}
\sum_{m=0}^{\delta}(-1)^{m}m{\delta \choose m}=0,
\label{linearalternating}
\end{equation}
regardless of the form of corresponding prefactors (which are functions of parameters $\alpha,k,l$).
It is left to check for special cases now -- at $k=l$ we calculate
\begin{eqnarray}
&(l-\alpha-1-x)(l-1-2x)H^{(\alpha)}_{l,l-1}(x)+(\alpha-l-x)(l-2x)H^{(\alpha+1)}_{l+1,l}(x)\nonumber \\
&+(\alpha-l-2x)(l-x)\left(H^{(\alpha+1)}_{l,l}(x)-H^{(\alpha)}_{l,l}(x)\right)\nonumber \\
&=\frac{1}{2}\Big[l(l-1-\alpha+x)(l-1-2x)f_{l-1}+(l+1)(\alpha-l-x)(l-2x)f_{l}\nonumber \\
&-(\alpha-l-2x)(l-x)f_{\alpha-l}\Big]=3l-\alpha,
\end{eqnarray}
and for $k=l-1$,
\begin{eqnarray}
&(\alpha-l+1-x)(l-1-2x)\left[H^{(\alpha+1)}_{l,l}(x)-H^{(\alpha)(x)}_{l-1,l-1}\right]\nonumber \\
&+(\alpha-l-2x)(-l+x)H^{(\alpha)}_{l-1,l}(x)\nonumber \\
&+(\alpha-l+1-2x)(l-1-x)H^{(\alpha+1)}_{l-1,l}(x)\nonumber \\
&=\frac{1}{2}\Big[(\alpha-l+1-x)(l-1-2x)f_{l-1}\nonumber \\
&-(\alpha-l+1)(\alpha-l-2x)(-l+x)f_{\alpha-l}\nonumber \\
&-(\alpha-l+2)(\alpha-l+1-2x)(l-1-x)f_{\alpha-l+1}\Big]\nonumber \\
&=3(l-1)-2\alpha.
\end{eqnarray}
One applies the same arguments to show the remaining case of the $E^{10}$ component.

\section{Nullspace vectors of $\bb{H}(x)$ and $\bb{H}(x)^{2}$}

\label{KerApp}

\paragraph*{1 .Vector $\bb{v}^{(\alpha)}$ is in the kernel of $\bb{H}^{(\alpha)}(x)$.}

It is sufficient to prove that
\begin{equation}
\bb{X}^{(\alpha)p}\bb{v}^{(\alpha)}=0,\qquad p=0,1,\ldots \alpha.
\end{equation}
This requirement is in fact implied by two separate (stronger, i.e. sufficient) conditions
\begin{equation}
\bb{Y}^{(\alpha)p}\bb{v}^{(\alpha)}=-\bb{v}^{(\alpha)},\qquad \mm{P}\bb{v}^{(\alpha)} =(-1)^\alpha \bb{v}^{(\alpha)}.
\end{equation}
The second being obviously satisfied, we focus on the first one and employ the component notation $\bb{v}^{(\alpha)}_{l}=(-1)^{l}$. We have to show that
\begin{equation}
\sum_{l=0}^{\alpha}(-1)^{k-p-1}{k \choose l}{k-l-1 \choose p-l}(-1)^{l}=(-1)^{k-1},
\end{equation}
which can be in turn recast into
\begin{equation}
\sum_{l=0}^{p}(-1)^{p+l}{k \choose l}{k-l-1 \choose p-l}=1.
\label{vsum1}
\end{equation}
One can quickly check that for $p=0$ the contribution comes only from $l=0$ and the equation trivially holds, which serves as our
\textit{basis of induction}. Next, by \textit{induction step} we move to $p\rightarrow p+1$, which after application of the Pascal's rule yields
\begin{eqnarray}
&\sum_{l=0}^{p+1}(-1)^{p+l}{k \choose l}\left[{k-l-1 \choose p-l}-{k-l \choose p-l+1}\right]\nonumber \\
&=\sum_{l=0}^{p}(-1)^{p+l}{k \choose l}{k-l-1 \choose p-l}-\sum_{l=0}^{p+1}(-1)^{p+l}{k \choose l}{k-l \choose p-l+1}.
\end{eqnarray}
On the right-hand side we retrieved an expression from the previous step plus an extra sum.
Introducing a summand function $V(\gamma,l)$, it is thus necessary to show that
\begin{equation}
\sum_{l=0}^{\gamma}V(\gamma,l):=\sum_{l=0}^{\gamma}(-1)^{l}{k \choose l}{k-l \choose \gamma-l}=0,\quad \gamma\geq 1,
\end{equation}
We rely on the observation that the sum at hand is \textit{Gosper-summable} \cite{petkovsek}, i.e. because the summand obeys the following \textit{recursive formula}
\begin{equation}
V(\gamma,l)=\Delta_{l}\left[\left(\frac{-l}{\gamma}\right)V(\gamma,l)\right],\quad \gamma\neq 0
\end{equation}
using the definition of the forward difference operator $\Delta_{m}A(m):=A(m+1)-A(m)$, the resulting \textit{telescoping series}
with finite support vanishes.

\paragraph*{2. Vector $\bb{u}^{(\alpha)}=(0,-1,2,-3,\ldots,(-1)^{\alpha}\alpha)^{T}$ is (i) an eigenvector of $\bb{Y}^{(\alpha)p}$ and
$\pi_{\rm a}(\bb{Y}^{(\alpha)p})$ with eigenvalue $-1$, provided $p\geq 1$. Additionally, (ii) for initial value $p=0$
we have $\bb{Y}^{(\alpha)0}\bb{u}^{(\alpha)}=0$ and $\pi_{\rm a}(\bb{Y}^{(\alpha)0})\bb{u}^{(\alpha)}=-\alpha\bb{v}^{(\alpha)}$.}
Therefore, (i) and (ii), together with $1.$ imply that $(\bb{H}^{(\alpha)})^{2}\bb{u}^{(\alpha)}=0$.

Initially for $p\geq 1$, rewriting the action of $\bb{Y}^{(\alpha)p}$ in components, and accounting for $\bb{u}^{(\alpha)}_l=(-1)^{l}l$, we obtain
\begin{equation}
\sum_{l=0}^{\alpha}Y^{(\alpha)p}_{k,l}u^{(\alpha)}_{l}=
\sum_{l=0}^{\alpha}(-1)^{k-p-1}{k \choose l}{k-l-1 \choose p-l}(-1)^{l}l=(-1)^{k}k.
\end{equation}
Since for $k=0$ it is evidently valid, we divide by $k$ and by means of ${k-1 \choose l-1}=\frac{l}{k}{k \choose l}$ and reformulate it as
\begin{equation}
\sum_{l=0}^{p}(-1)^{p}F(p,l):=\sum_{l=0}^{p}(-1)^{l-p}{k-1 \choose l-1}{k-l-1 \choose p-l}=1\quad, k\geq 1.
\end{equation}
Beginning with the basis of induction at $p=1$, we first show
\begin{equation}
\sum_{l=0}^{1}-F(p,l)=-F(1,1)=1.
\end{equation}
Proceeding with the inductive step $p\rightarrow p+1$ we find
\begin{eqnarray}
\sum_{l=0}^{p+1}F(p+1,l)&=\sum_{l=0}^{p+1}(-1)^{l-p+1}{k-1 \choose l-1}{k-l-1 \choose p-l+1}\nonumber \\
&=\sum_{l=0}^{p+1}(-1)^{l-p}{k-1 \choose l-1}\Bigg[{k-l-1 \choose p-l}-{k-l \choose p-l+1}\Bigg]\nonumber \\
&=\sum_{l=0}^{p}F(p,l)-\sum_{l=0}^{p+1}(-1)^{l-p}{k-1 \choose l-1}{k-l \choose p-l+1}.
\end{eqnarray}
It is necessary to show, that the second sum always vanishes for $\gamma \geq 2$,
\begin{equation}
\sum_{l=0}^{\gamma}\tilde{V}(\gamma,l):=\sum_{l=0}^{\gamma}(-1)^{l}{k-1 \choose l-1}{k-l \choose \gamma-l}=0,
\end{equation}
which is again summed up by help of the recursive formula for the summand,
\begin{equation}
\tilde{V}(\gamma,l)=\Delta_l\Big[\frac{l-1}{1-\gamma}\tilde{V}(\gamma,l)\Big].
\end{equation}

Finally, we prove exceptional cases at $p=0$, where only $l=0$ contributes, and consequently
\begin{eqnarray}
(\bb{Y}^{(\alpha)0}\bb{u}^{(\alpha)})_k &=Y^{(\alpha)0}_{k,0} u^{(\alpha)}_{0}=0,\nonumber \\
\left(\pi_{\rm a}(\bb{Y}^{(\alpha)0})\bb{u}^{(\alpha)}\right)_k &=Y^{(\alpha)0}_{\alpha-k,0}u^{(\alpha)}_{\alpha}=(-1)^{k}(-\alpha)=-\alpha v^{(\alpha)}_{k}.
\end{eqnarray}
Hence, $\pi_{\rm a}(\bb{Y}^{(\alpha)0})\bb{u}^{(\alpha)}=-\alpha \bb{v}^{(\alpha)}$ and consequently also
\begin{equation}
\bb{X}^{(\alpha)0}\bb{u}^{(\alpha)}= \frac{\alpha}{2} \bb{v}^{(\alpha)}.
\end{equation}
By combining the above results with the residue form of $\bb{H}^{(\alpha)}(x)$ (\ref{residueform}) we conclude
\begin{equation}
\bb{H}^{(\alpha)}(x)\bb{u}^{(\alpha)}=\frac{\alpha}{x}\bb{v}^{(\alpha)}.
\end{equation}

\end{document}